\documentclass[useAMS,twocolumn,usenatbib]{mn2e}
\pdfoutput=1
\usepackage{times,psfig,epsfig}
\usepackage{rotating, graphicx}
\usepackage{color}
\usepackage{amssymb, amsmath}
\usepackage{multirow}
\usepackage{subfigure}

\usepackage[titletoc]{appendix}

\usepackage{enumerate}
\newif\ifAMStwofonts
\AMStwofontstrue

\def\rvir{R_{\rm vir}}

\def\rfive{R_{\rm 500}}

\def\rtwo{R_{\rm 200}}

\def\msunh{\,h^{-1}\rm{\,M_{\odot}}}

\def\kms{\rm{\,km\,s^{-1}}}

\def\agn{{\tt AGN}}
\def\csf{{\tt CSF}}

\newcommand{\be}{\begin{equation}}
\newcommand{\ee}{\end{equation}}
\def\aap{A\&A}

\def\apjl{ApJ}

\def\apjs{ApJS}

\newcommand{\chandra}{{\it Chandra}}
\newcommand{\asca}{{\it ASCA}}
\newcommand{\sax}{{\it BeppoSAX}}

\newcommand{\xmm}{{\it XMM-Newton}}

\newcommand{\suzaku}{{\it SUZAKU}}
\newcommand{\athena}{{\it Athena}}

\title[Metal enrichment of the ICM]{The history of chemical enrichment in the intracluster medium from cosmological simulations}

\author[V.~Biffi et al.]{V.~Biffi$^{1,2}$\thanks{e-mail: biffi@oats.inaf.it}, S.~Planelles$^{3}$, S.~Borgani$^{1,2,4}$, D.~Fabjan$^{5,2}$, E.~Rasia$^{2,6}$, G.~Murante$^2$,
\newauthor
L.~Tornatore$^2$, K.~Dolag$^{7,8}$,
G.L.~Granato$^2$, M.~Gaspari$^9$, A.M.~Beck$^7$
\\~\\
\footnotesize
$^1$ Dipartimento di Fisica dell' Universit\`a di Trieste, Sezione di Astronomia, via Tiepolo 11, I-34131 Trieste, Italy\\
$^2$ INAF, Osservatorio Astronomico di Trieste, via Tiepolo 11, I-34131, Trieste, Italy\\
$^3$ Departamento de Astronom{\'i}a y Astrof{\'i}sica, Universidad de Valencia, c/ Dr. Moliner, 50, 46100 - Burjassot (Valencia), Spain\\
$^4$ INFN --- National Institute for Nuclear Physics, Via Valerio 2, I-34127 Trieste, Italy\\
$^5$ Faculty of Mathematics and Physics, University of Ljubljana, Jadranska 19, 1000 Ljubljana, Slovenia \\
$^6$ Department of Physics, University of Michigan, 450 Church St., Ann Arbor, MI 48109, USA\\
$^7$ University Observatory Munich, Scheinerstr. 1, D-81679 Munich, Germany\\
$^8$ Max-Planck-Institut f\"ur Astrophysik, Karl-Schwarzschild Strasse 1, 85748 Garching bei M¬unchen, Germany\\
$^9$ Department of Astrophysical Sciences, Princeton University, Princeton, NJ 08544, USA; Einstein and Spitzer Fellow
}

\begin{document}
\maketitle
\begin{abstract}
The distribution of metals in the intracluster medium (ICM) of galaxy
clusters provides valuable information on their formation and evolution,
on the connection with the cosmic star formation and
on the effects of different gas processes.  By
analyzing a sample of simulated galaxy clusters, we study the chemical
enrichment of the ICM, its evolution, and its relation with the
physical processes included in the simulation and with the thermal
properties of the core.  These simulations, consisting of
re-simulations of 29 Lagrangian regions performed with an upgraded
version of the SPH {\footnotesize GADGET-3} code, have been run including two
different sets of baryonic physics: one accounts for radiative
cooling, star formation, metal enrichment and supernova (SN) feedback,
and the other one further includes the effects of feedback from active
galactic nuclei (AGN).  In agreement with observations, we find an
anti-correlation between entropy and metallicity in cluster cores, and
similar radial distributions of heavy-element abundances and
abundance ratios out to large cluster-centric distances ($\sim R_{180}$).  In
the outskirts, namely outside of $\sim 0.2\,R_{180}$, we find a remarkably
homogeneous metallicity distribution, with almost flat profiles of the
elements produced by either SNIa or SNII.  We investigated the origin
of this phenomenon and discovered that it is due to
the widespread displacement of metal-rich gas by early ($z>2$--$3$) AGN
powerful bursts, acting on small high-redshift haloes.
Our results also indicate that the intrinsic metallicity of the hot gas
for this sample is on average consistent with no evolution between $z=2$ and $z=0$, across the entire radial range.
\end{abstract}

\begin{keywords}
{galaxies: clusters: general --- galaxies: clusters: intracluster medium --- methods: numerical}
\end{keywords}

\section{Introduction}

In the hierarchical model of structure formation, galaxy clusters, the
youngest and largest cosmic structures to form, emerge from the
gravitational collapse of large overdense regions in the matter
distribution.  Although most part of the cluster mass (typically,
$10^{14}$--$10^{15} {\rm M_{\sun}}$) is in the form of dark matter ($\sim$
85\%), whose dynamics dominates the cluster formation, there is also a
contribution from stars in galaxies ($\sim$ 5\%) and from a hot
($T\sim 10^7$--$10^8$\,K) and diffuse plasma, the intracluster medium
(ICM; $\sim 10$--$15$\%), which can be observed through its X-ray emission
\citep[see][for reviews]{kravtsov2012, Planelles_2016}.  Given their
large masses, clusters are characterized by very deep gravitational
potential wells which allow us to study in detail all the baryonic
processes affecting both the stellar component and the hot ICM
simultaneously.

\looseness=-1
X-ray spectral observations of galaxy clusters~\citep[see][for a
  review]{Bohringer_2010} provide important information on the
ICM temperature and density, and on the abundance of different
metals (chemical elements heavier than helium).
These metals are mainly released during the life phases
  of low- and intermediate-mass stars within galaxies and by supernova
  (SN) explosions.
In particular, whereas core-collapse or Type II supernovae (SNII) are mainly responsible for the production of light metals (such as, O, Ne, Mg or Si --- but contribute also to Fe enrichment),  Type Ia supernovae (SNIa) produce large amounts of heavier metals (principally, Fe and Ni). Intermediate-mass elements, such as, Si, S, Ar or Ca, are produced by both SNe types in similar proportions, whereas lighter elements (C, N) are contributed by low- and intermediate-mass stars
during their AGB (Asymptotic Giant Branch) phase.
In addition to the different metal production, SNII and SNIa also show different timescales in releasing the metals. In fact, while the former result from high-mass stars, whose lives are short and trace therefore recent star formation episodes, the latter result from long-lived low-mass stars and can thus be delayed by billions of years from the time of star formation, contributing to the late enrichment.
Therefore,  observations of the relative abundance ratios  of different elements, and their spatial distributions within the cluster volume, provide valuable information on the number and distribution of each SN type and on their metal production efficiency throughout the cluster evolution.
Moreover, the fact that the ICM is metal-rich throughout the cluster volume (and not only in the centre of star forming galaxies) suggests that there should be an interplay between  a number of astrophysical and dynamical processes, such as galactic winds, AGN feedback or ram-pressure stripping, which contribute to mix and redistribute the metal content of the ICM~\citep[e.g.][]{Churazov_2001, Rebusco_2005, Rebusco_2006, Simionescu_2008, Simionescu_2009, Gaspari_2011a, Gaspari_2011b, Kirkpatrick_2015}.

The first observations with enough sensitivity to study the ICM metal content were provided by the X-ray satellites \asca\   and \sax.  Early \asca\ observations confirmed an average Fe enrichment in the ICM of around $\sim1/3$ of the solar abundance and provided, for the first time, abundance measurements for other elements~\citep[e.g.][]{Mushotzky_1996, Finoguenov_2000}.
On the other hand, early observations with \sax\ of the radial distribution of the ICM Fe abundance in local massive galaxy clusters showed a significant trend: while non-cool core (NCC) clusters appear to have an almost flat iron abundance profile out to $\sim 0.4 R_{180}$\footnote{$R_{180}$ is defined as the radius of the sphere enclosing the volume of the cluster whose mean density equals $180$ times the critical density of the Universe. In a similar way, we will also define $\rtwo$ and $\rfive$, used in the following.}, cool-core  (CC) systems show instead a significant enhancement of the Fe abundance in their cores
\citep[][]{deGrandiMolendi2001, deGrandi_2004, Bohringer_2004, baldi2007}.

Deeper and better resolved observations of galaxy clusters with \xmm, \chandra\ and \suzaku\  allowed for  a more detailed analysis of the ICM metal content and its spatial distribution. In local clusters, the level of core ($r<0.1R_{180}$) entropy anticorrelates with the central enrichment level~\citep{leccardi2010}. In addition, the analysis of the Si/Fe radial distribution reveals a nearly flat  profile within the central cluster regions, suggesting a flat proportion and a similar metal contribution of SNII and SNIa~\citep[e.g][]{sato2008, sakuma2011, matsushita2013}. These X-ray observations were still limited by a relatively poor spatial resolution and, as a consequence, most of the data provided correspond to central cluster regions.
More recently, observations with \suzaku\ of the Perseus and the Virgo clusters have provided the first observational determination of the metal abundance out to cluster outskirts~\citep{werner2013, simionescu2015}, confirming a relatively uniform Fe distribution throughout the cluster volume. To explain such an extended metal distribution,  an early ($z>2$) ICM enrichment, is required.

A number of studies have also addressed  the z-evolution of the ICM metal abundance, suggesting, however, contradictory trends~\cite[e.g.][]{Balestra_2007, Maughan_2008, Baldi_2012, ettori2015, McDonald2016}. Recently, cluster observations by~\cite{McDonald2016}, out to $z\sim 1.5$,  report a weak z-evolution of the central metallicity in CC clusters, although this correlation seems to be less strong than the one reported by~\cite{ettori2015}. Instead, when the core is excised, either works do not find a significant difference between CC and NCC clusters and both populations seem to show a non-evolving metal abundance in their outer regions~\cite[see][for a critical analysis of the difficulties in assessing observationally the ICM metal content]{Molendi_2016}.

Within this context, cosmological hydrodynamical simulations of galaxy clusters are extremely useful to shed some light on the complex processes affecting the  ICM metal distribution, its  z-evolution, its connection with the metal production by different SN types, its relation with existing dynamical and feedback processes and the possible biases on the metallicity measures~\citep[e.g.][]{tornatore2004, tornatore2007, Cora_2008, rasia2008, Fabjan_2008, Fabjan_2010, mccarthy2010, planelles2014, Martizzi_2016}. Any hydrodynamical simulation that tries to describe, consistently with the cosmic evolution, the production of heavy elements must couple complicated models of chemical enrichment  with a number of subgrid physical processes, such as star formation, SN feedback, galactic winds or AGN feedback.  These chemical evolution models (characterized by a number of elements, such as the SNe explosion rates and lifetime functions, the stellar yields or the initial mass function) are crucial to obtain a global picture of the cosmic enrichment. Indeed, despite the inevitable uncertainties associated to these models, cosmological hydrodynamical simulations, including different chemical enrichment schemes, have confirmed many of the observations of the local metal content in galaxy clusters~\citep[see][for a review]{Borgani_2008}.

In this paper, we take advantage of the simulated sample of galaxy clusters recently presented by~\cite{rasia2015} that include the effects of radiative cooling, star formation, chemical enrichment and AGN feedback.
These simulations, for the first time, reproduce the observed ratio of CC and NCC clusters and are in good agreement with observations, in terms of iron abundance and entropy profiles.
Starting from this result, we perform, here, a detailed analysis of the ICM metal  distribution. In particular,  the abundances, spatial distributions and ratios of different elements, such as Fe, Si and O, will be analyzed as a function of the different physics included in the simulations, the cool-coreness of the systems and the z-evolution. The production of metals by different sources tracked in the simulations, namely, SNIa, SNII and AGB stars, will be also discussed.

The organization of the paper is as follows. A brief description of
our simulations, together with the main properties of the sample of
clusters to be analyzed, is introduced in Section
\ref{sec:simulations}. Results on the chemical properties of our set
of simulated galaxy clusters at z=0 are shown in Section
\ref{sec:met}, whereas the z-evolution of the ICM chemical enrichment is
analyzed in Section \ref{sec:evol-met}. In Section~\ref{sec:AGN-CSF}
we discuss the effects of AGN feedback on the distribution of heavy
elements in the ICM, and in Section \ref{sec:track} we present the
different contributions to the ICM metal content by three different
enrichment sources. Finally, in Section \ref{sec:conclusions} we
discuss and summarize our main findings.

\section{Numerical simulations}
\label{sec:simulations}

The numerical simulations analyzed in this paper have been already
presented in a number of recent works~\cite[][]{rasia2015,villaescusa2016,Biffi_2016,Truong_2016,planelles2017}.
  Therefore, here we only detail the most relevant features for the current study.  For
further numerical or technical information on this set of simulations, we
refer the reader to these papers.

Our set of simulations consists in 29 zoomed-in Lagrangian regions
simulated with an upgraded version of the {\footnotesize GADGET-3}
code~\cite[][]{springel2005}.  These regions, selected from a parent
DM-only simulation, were extracted around 24 dark matter
halos with $M_{200}>8\times 10^{14}\, \msunh$ and 5 smaller ones
with $M_{200}=[1-4]\times 10^{14}\, \msunh$~\cite[see][for more details]{bonafede2011}.  Each low-resolution
region was then re-simulated to achieve better resolution and include the
contribution from baryons.  The simulations have been run with these
cosmological parameters: $\Omega_{\rm{m}} = 0.24$, $\Omega_{\rm{b}} =
0.04$, $n_{\rm{s}}=0.96$, $\sigma_8 =0.8$ and
$H_0=72\,\kms$\,Mpc$^{-1}$.

The mass resolution for the DM and gas particles is
$m_{\rm{DM}} = 8.47\times10^8 \, \msunh$ and $m_{\rm{gas}} =
1.53\times10^8\, \msunh$, respectively.  For the gravitational force, a
Plummer-equivalent softening length of $\epsilon = 3.75\, h^{-1}$\,kpc
is used for DM and gas particles, whereas $\epsilon = 2\, h^{-1}$\,kpc
for black hole and star particles.  The DM
  softening length is kept fixed in comoving coordinates for $z>2$ and
  in physical coordinates at lower redshift. For the other
  components, the softening lengths are always determined in comoving
  coordinates.

As for the hydrodynamical scheme, the improved SPH formulation,
described by~\cite{beck2015}, was implemented. This new SPH model,
that includes artificial thermal diffusion and higher-order interpolation kernel,
improves the standard SPH performance in
its capability of treating discontinuities and following the development
of gas-dynamical instabilities.

Our set of simulations includes different treatments of
baryonic physics.  In particular, in the following we will analyze two
of~them:
\begin{itemize}
\item {\tt CSF}. These simulations include radiative cooling, star
  formation, SN feedback and metal enrichment, as introduced in~\citet{planelles2014}. Briefly,  star formation and its associated feedback
  are implemented according to the original model by~\cite{springel2003},
the heating/cooling from Cosmic Microwave Background (CMB) and
from a UV/X--ray time-dependent uniform ionizing background  are included as in~\cite{haardt_madau01}, and the rates of radiative cooling
follow the prescription by~\cite{wiersma_etal09}.

\item {\tt AGN}. These runs include the same physical
  processes as in the \csf{} case, but account as well for AGN
  feedback.  Our model for SMBH accretion and AGN feedback
  was presented in~\citet{steinborn2015}.

\end{itemize}

\subsection{The model for chemical enrichment}\label{sec:chem-mod}
Stellar evolution and metal enrichment are included following the
original formulation by~\cite{tornatore2007}.

Our model of star formation is an updated version of the original
implementation by~\cite{springel2003}. In particular, gas particles
with a density above $0.1\,$cm$^{-3}$ and a temperature below
$2.5\times 10^5$\,K are classified as multiphase, with a cold and
a hot-phase coexisting in pressure equilibrium. The cold phase represents
the reservoir for star formation.

The production of heavy elements
considers the contributions from SNIa, SNII and AGB stars.  Whereas all three types of stars contribute to the
chemical enrichment, only SNIa and SNII provide as well
thermal feedback.  In addition, as described in~\cite{springel2003},
kinetic feedback from SNII is implemented as galactic outflows with a
wind velocity of $350 \kms$.
The distribution of initial masses for the population of stars is
provided by the initial mass function of~\cite{chabrier03}.
We assume the mass-dependent lifetimes of~\cite{padovani_matteucci93}
to account for the different time-scales of stars of different masses.

To estimate the abundance of the different metal species produced
during the evolution of a stellar particle, we consider different sets
of stellar yields.  In particular, we have adopted the yields provided
by~\cite{Thielemann2007} for SNIa stars and those by
\cite{karakas2007} for AGB stars.  In the case of
SNII, we use the metal-dependent yields by~\cite{WoosleyWeaver1995}
combined with those by~\cite{romano2010}.
In our simulations, we specifically trace the
production and evolution of 15 chemical elements: H, He, C, Ca, O, N, Ne,
Mg, S, Si, Fe, Na, Al, Ar and Ni.
Metallicity-dependent radiative cooling is included in our
simulations following the approach proposed in~\cite{wiersma_etal09}, and
considering the contribution of the fifteen chemical species traced by
using pre-computed tables from the {\tt CLOUDY} code for an optically
thin gas in (photo-)ionization equilibrium.

Given the features of the chemical evolution model implemented in our
simulations, it is relatively straightforward to estimate the
contribution to metallicity by a particular stellar source. Therefore,
at any given redshift, we can compute the amount of the mass in metals
in a gas or stellar particle provided by either SNIa, SNII or AGB
stars.  For further details on the chemical evolution model, we refer
the interested reader to~\cite{tornatore2007}.

Despite the large variety of physical processes treated, our
simulations still present some limitations.  Our chemical model, for
instance, does not include any treatment of metal diffusion, although
the code accounts for the spreading of metals from star particles to
the surrounding gas particles, by using the same kernel of the SPH
scheme.  This is mainly done in order to avoid a noisy estimation of
metal-dependent cooling rates, and thus heavy elements can be
spatially distributed only via dynamical processes involving the
metal-rich gas.  Also, another important process that is not yet
accounted for in our simulations is the metal depletion due to the
formation of dust grains. Since a possibly large fraction of heavy
elements (especially C, O, S, Fe, Mg, Si) are likely to be locked into
dust, solid particles, this could also affect the distribution of
gas-phase metallicity.  For a comprehensive study of the enrichment
level in clusters, in which a full picture of the heavy-elements cycle
among the cluster baryonic components (stars, multi-phase ICM and
dust) can be drawn, these effects should be ultimately taken into
account.

\subsection{The set of simulated clusters}

For the purpose of the present analysis, we concentrate on the 29 main
halos residing in each of the re-simulated Lagrangian regions.
The sample consists of 29 clusters
with masses $6\times
10^{13}h^{-1}{\rm M_\odot}  \leq M_{500} \leq 2\times
10^{15}h^{-1}{\rm M_\odot}$
at $z=0$~\cite[e.g.][]{planelles2014}.
Their centre is found by applying a Friends-of-Friends (FoF) analysis to
identify the position of the particle with the minimum of the
gravitational potential within the FoF group. Given this, we determine
the cluster virial radius $\rvir$ by adopting a spherical collapse
algorithm, based on this centre. $\rvir$ is defined as the radius
encompassing an average overdensity, $\Delta_{\rm vir}$, which is
equal to the overdensity predicted by the spherical collapse
model~\cite[e.g.][]{bryan1998}, in units of the critical density of the
Universe $\rho_c(z)=3H(z)/(8\pi G)$.
In addition to $\rvir$, we will also make use of radii defined for
$\Delta=500,200,180$,\footnote{For the clusters in our sample,
  $\rfive:\rtwo:R_{180}:\rvir$ is on average $0.61:0.95:1:1.34$, at
  $z=0$.} in order to either ease the comparison to particular
observational datasets or to investigate the predicted properties of
the simulated clusters within interesting regions.

As presented in~\cite{rasia2015}, our \agn{} simulations reproduce the
diversity of core thermodynamical properties observed in real
clusters, with entropy and metallicity profiles that resemble quite
well the differences between observed cool-core (CC) and non-cool-core
(NCC) populations.  These two classes were defined on the basis of
observational criteria.  In particular, we define CC cluster a system
with a central entropy\footnote{We adopt here the definition of ``entropy'', K,
  typically used in X-ray studies of galaxy clusters, that is $K=k_B T /
  n_e^{2/3}$, where $T$ is the temperature and $n_e$ is the electron
  density of the gas.}  $K_0< 60\,$keV\,cm$^{2}$ and a pseudo-entropy
$\sigma<0.55$ (see Sec.~\ref{sec:entr-met} for the definition of the
latter); we classify a cluster as a NCC when it does not satisfy both
aforementioned conditions.  With this classification, our sample of
simulated clusters is sub-divided into 11~CC and 18~NCC.
That paper demonstrates that the low isentropic level reached by the
CC systems is achieved thanks to the efficient AGN feedback that
contrasts the radiative cooling.

\begin{figure*}
\centering
{\bf (a) CC cluster} \hspace{210pt} {\bf (b) NCC cluster}\\[7pt]
\setlength{\tabcolsep}{1pt}
\begin{tabular}{ccccc}
\includegraphics[width=0.23\textwidth]{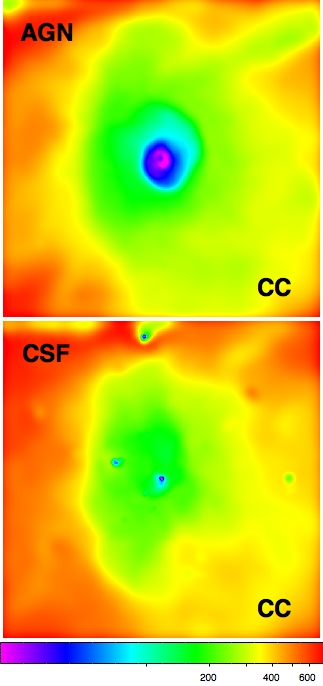}&
\includegraphics[width=0.23\textwidth]{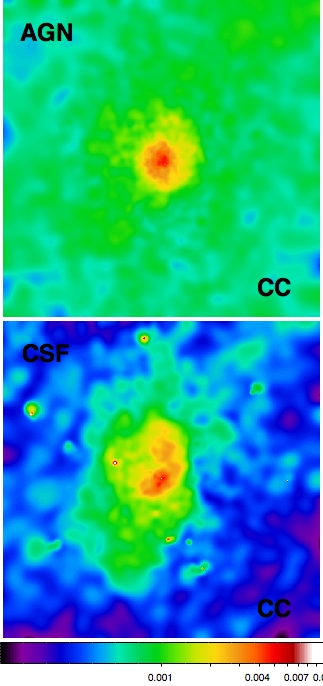}
&\hphantom{blahblah}&
\includegraphics[width=0.23\textwidth]{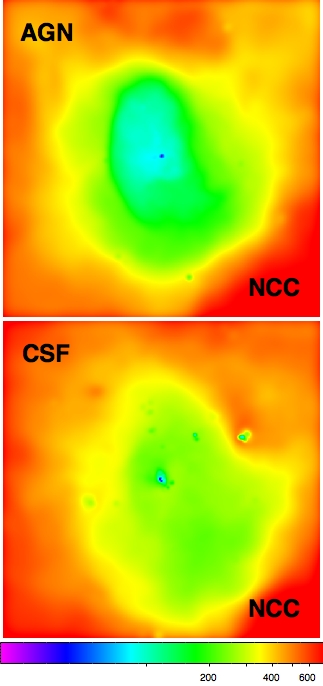}&
\includegraphics[width=0.23\textwidth]{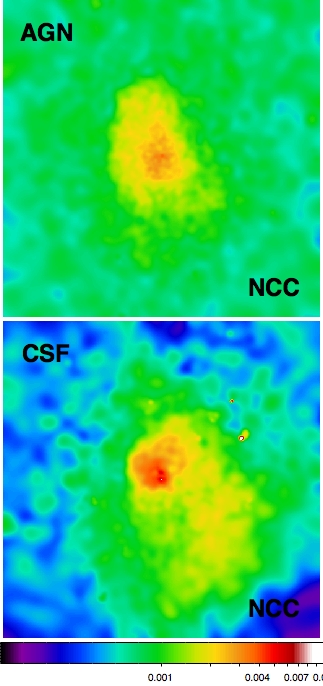}
\\[5pt]
pseudo-entropy & oxygen abundance && pseudo-entropy & oxygen abundance
\end{tabular}\\[7pt]
\caption{Projected maps of pseudo-entropy (left) and oxygen abundance
  (right) for a CC (a) and a NCC (b) cluster in the sample, at
  $z=0$. For each cluster we show both the result from the \agn{} run
  (top panels) and from the \csf{} one (bottom panels). The
  pseudo-entropy maps are derived from the maps of spectroscopic-like
  temperature and X-ray soft band ([0.5--2]\,keV) emissivity combined
  as ${\cal M}_{\rm TSL}/{\cal M}_{\rm SXR}^{1/3}$.  The oxygen
  abundance is reported as mass fraction with respect to hydrogen.
  Each map is centred on the position of the minimum of the cluster
  potential well, and comprises a region of $2$\,Mpc per side and
  $5$\,Mpc of integration along the line of sight.}\label{fig:maps}
\end{figure*}

The effect produced by the AGN action
on the thermodynamical and metallicity properties can be appreciated
in Fig.~\ref{fig:maps} where we compare
two clusters classified as CC and NCC in the \agn{} run
(Figs.~\ref{fig:maps}(a) and~\ref{fig:maps}(b), respectively), against
their \csf{} counterparts (lower panels).  In this way, the different
central thermal and chemical properties can be appreciated.

For each cluster, we show the pseudo-entropy and metallicity maps of
the ICM at $z=0$.  The pseudo-entropy maps are created via simple
arithmetics with the maps of the spectroscopic-like temperature,
${\cal M}_{\rm TSL}$, and of the emissivity, ${\cal M}_{\rm SXR}$,
in the soft X-ray band ($[0.5$--$2]\,$keV), as ${\cal M}_{\rm
  TSL}/{\cal M}_{\rm SXR}^{1/3}$~\cite[see e.g.][]{finoguenov2010}.
In order to visualize the ICM metallicity, we show the
    oxygen abundance.
  The maps are centred on each cluster centre and cover a volume of
  $2$\,Mpc per side, thus enclosing the projected $\rfive$ in both
  cases. The projection is done for $5$\,Mpc along the line of sight.
The 2D maps employed are generated by using the {\sc
      Smac} program~\cite[][]{dolag2005}, designed to integrate
    various physical and observational-like quantities (in this case,
    spectroscopic-like temperature, soft X-ray emissivity and oxygen
    abundance) through the simulation box at any given redshift.

  In general, we note that both clusters appear as NCC systems in the
  \csf{} case, for what concerns the level of central
  (pseudo)~entropy, which is significantly higher than in their \agn{}
  counterparts.  In particular, the cool core developed in the central
  region of the CC pseudo-entropy map (top-left panel in
  Fig.~\ref{fig:maps}(a)) does not appear in the corresponding \csf{}
  map.  Also, from the oxygen abundance maps, we notice that the
  difference in the level of metal enrichment in the core of the two
  \agn{} clusters, higher and more concentrated in the CC system than
  in the NCC one (top-right panels in the two figures), is not
  observed in the \csf{} case, where the metallicity in the core is
  high for both clusters (differently from observed NCC systems).
  The distribution of the oxygen-rich gas in the two runs, both for
  the CC and the NCC cluster, also shows remarkable large-scale
  differences between the \agn{} and the \csf{} run. In the first case
  (top panels) the distribution appears to be more homgeneous, with a
  generally higher value of $Z_{O}$ out to $\sim 1$\,Mpc from the
  cluster centre. Instead, the \csf{} clusters present a more clumpy
  distribution of the oxygen, with high-metallicity small-scale
  substructures, well outside the central $\sim (250\,{\rm kpc})^2$
  (enclosing approximately $\sim 0.25\,\rfive$ for both
  clusters). From the visual comparison of the \agn{} and the \csf{}
  oxygen maps we can already infer that the gradient of the ICM
  metallicity from the cluster centre to the outskirts is steeper in
  the \csf{} clusters.


\section{Metal enrichment of the ICM}\label{sec:met}
Here, we present the chemical properties of the 29 simulated
 galaxy clusters, at redshift $z=0$.    These
 results are also compared to available observations of clusters in the
 local Universe.

 For the purpose of a fair comparison to the X-ray observational data,
 we will consider projected quantities weighted by the emissivity of
 the gas. These will include a large volume of the cluster due to the
 integration along the line of sight.

The intrinsic metal content of the ICM will instead be evaluated by
computing the three-dimensional mass-weighted values, for both global
quantities and radial profiles.

 Throughout the paper, the reference solar values are assumed
 according to the abundance pattern reported
 by~\cite{angr1989}\footnote{Specifically, the solar abundance values
   we use are: $4.68\times 10^{-5}$ for Fe, $3.55\times 10^{-5}$ for
   Si and $8.51\times 10^{-4}$ for O (number fractions, relative to
   hydrogen).}. Observational results are therefore rescaled to these, when
 different.

 \subsection{The entropy-metallicity relation}\label{sec:entr-met}
\begin{figure}
\includegraphics[width=0.45\textwidth,trim=30 7 20 15,clip]{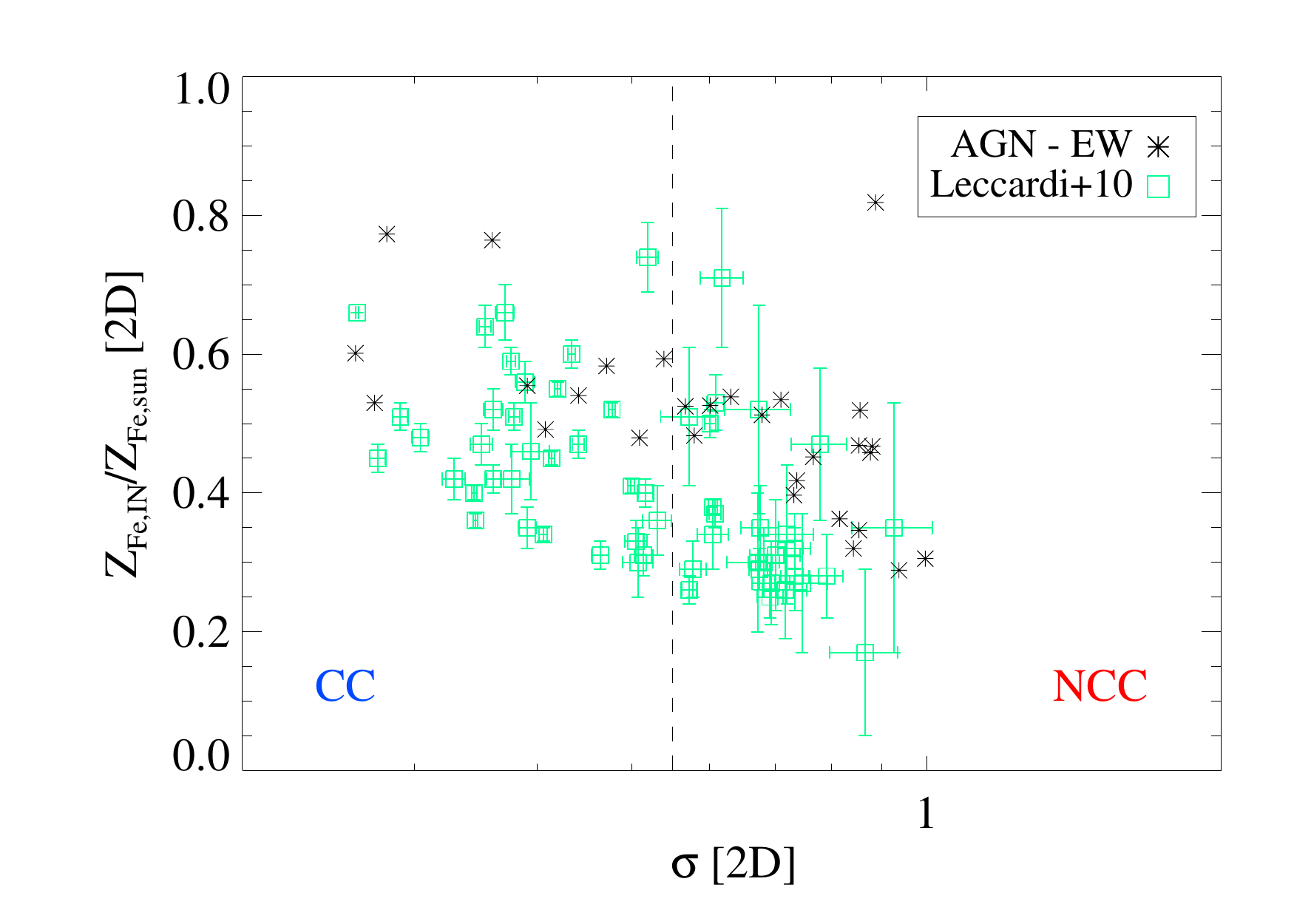}
\includegraphics[width=0.45\textwidth,trim=30 0 20 5,clip]{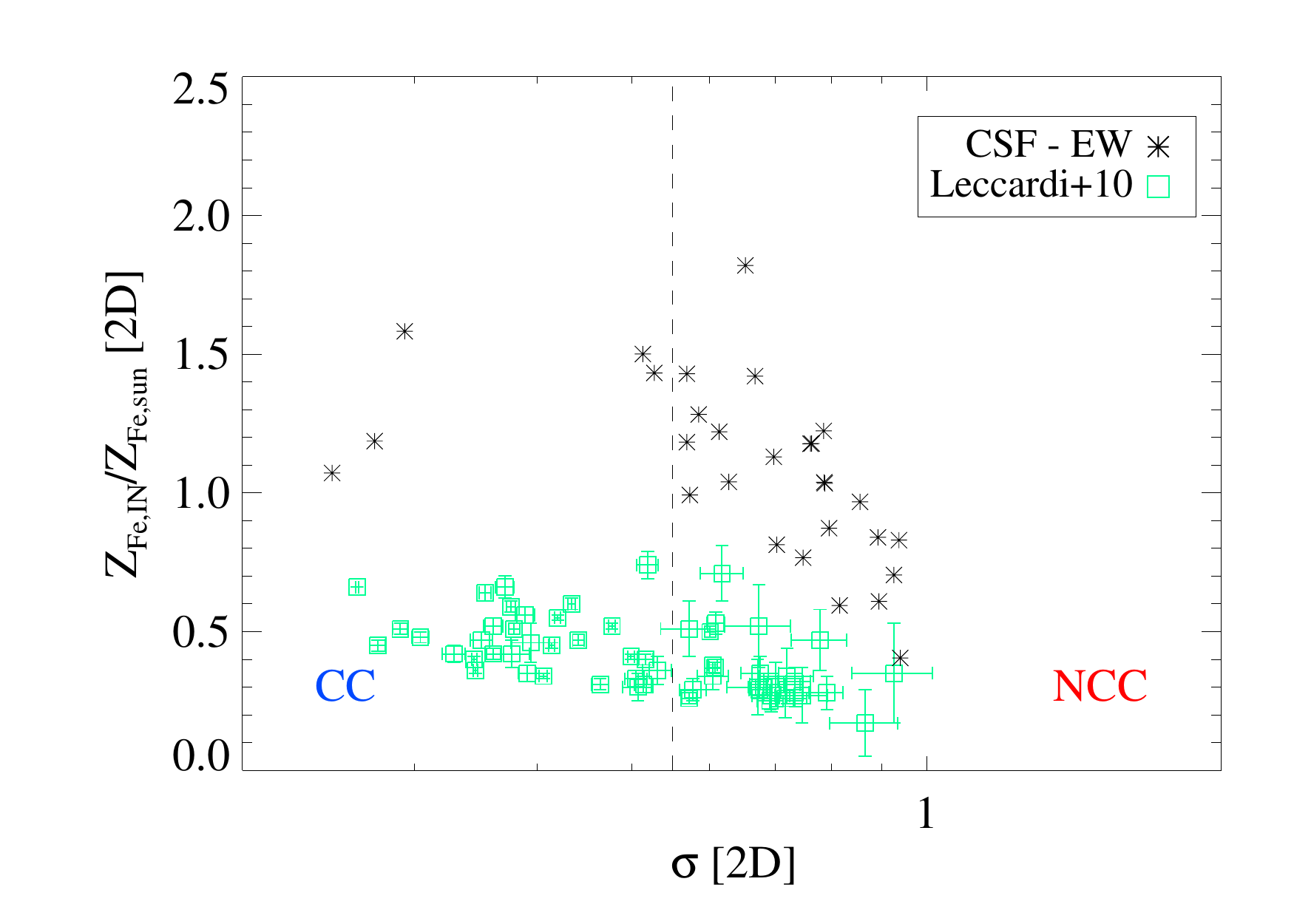}
 \caption{Relation between emission-weighted metallicity and
   pseudo-entropy ratio at $z=0$, for our sample of clusters (black
   asterisks).  The upper panel shows projected (2D) simulation data
   for the \agn{} run, whereas the lower panel displays the same
   quantities for the \csf{} set. In both cases, we show observational
   data from~\protect\cite{leccardi2010}, for comparison (green
   squares with error bars).  The vertical dashed line at
   $\sigma=0.55$ marks the threshold separating between CC and NCC
   (originally applied to 3D values of $\sigma$
   in~\protect\cite{rasia2015}). Notice the different $y$-axis~range.}
\label{fig:entr-met0}
\end{figure}

As a first test on the connection between thermodynamical and chemical
properties of the simulated clusters in the \agn{} run, we explore the
relation between the pseudo-entropy ratio $\sigma$ and the central
metallicity, at $z=0$. This is shown in Fig.~\ref{fig:entr-met0}.

The pseudo-entropy ratio is defined as $\sigma = (T_{\rm IN}/T_{\rm OUT}) * ({\rm
  EM}_{\rm IN}/{\rm EM}_{\rm OUT})^{-1/3}$, with the
spectroscopic-like temperature (T) and emission measure (EM) computed
within the projected ``IN'' and ``OUT'' regions, corresponding to
$r<0.05\,R_{\rm 180}$ and $0.05\,R_{\rm 180} < r < 0.2\,R_{\rm 180}$,
respectively~\cite[e.g.][]{leccardi2010,rossetti2011}.
The metallicity is computed as the emission-weighted iron abundance
within the projected ``IN'' region.

In Fig.~\ref{fig:entr-met0}, we compare the simulation
predictions to observational findings by~\cite{leccardi2010}, shown in green.
To this scope, the
projected values
of both metallicity and
pseudo-entropy are computed in cylindrical regions
of $2\,\rvir$-length along the line of sight.
In general, the differences in normalization can be related to several
factors. For instance, the yields considered in the numerical
modelling, the assumed initial mass function within the simulations,
as well as the lack of a treatment of proper metal diffusion
and dust depletion, can affect the level of chemical enrichment of the ICM.

We recover the observed anti-correlation between $Z_{\rm
  Fe,IN}$ and $\sigma$, namely the lower the pseudo-entropy, the stronger the metallicity peak.
Specifically, the best fit\footnote{We follow a Bayesian approach to
  linear regression, and perform the fit by using the {\small IDL} macro
  {\tt linmix\_err.pro} by B.C.~Kelly~\cite[][]{kelly2007}.}  to the
anti-correlation for the simulated data provides a normalization of
$A_{\rm sim} = 0.76 \pm 0.05$ and a slope of $\beta_{\rm sim} =
-0.41\pm 0.07$, respectively.
Compared to the observational data by~\cite{leccardi2010},
we therefore find a $\sim 15\%$ higher, albeit consistent at
1-$\sigma$, normalization (w.r.t. $A_{\rm obs} = 0.66 \pm 0.05$, for
the observed relation). Concerning the slope, our result is instead
fully consistent with the observed value of $\beta_{\rm obs} =
-0.47\pm 0.10$~\footnote{We fit both simulated and observed data by
  using the observational errors for the latter, and cosindering a
  constant error value, equal to 0.01, in both variables in the former
  case.}.  The good agreement in slope confirms that our \agn{}
simulations are capable to obtain metal-rich gas that also tends to be
the low-entropy gas.  The intrinsic scatter is relatively similar,
that is $\sim 8\%$ for the simulations and $\sim 10 \%$ in the
observations.  To obtain these results we exclude the data point
corresponding to the outlier in the top-right region of the
plot. Through a deeper investigation, this cluster appears to be a
very small mass system with a significant substructure interacting
with the main halo. During its redshift evolution, this kind of
interactions have affected the build-up of a cool core without
removing however the highly enriched gas from its central regions.

For comparison, we show in the lower panel of Fig.~\ref{fig:entr-met0}
the entropy-metallicity relation, at $z=0$, for the \csf{} sample of
simulated clusters.  (Notice that the maximum value shown in the y-axis
has been changed to accomodate the \csf{} results.)
In this case, the slope is found to be excessively steep in
comparison to the one observed.
From the distribution of the data points we note that there is a
significant enhancement of the median emission-weighted Fe abundance
in the core region, from $Z_{\rm Fe,IN}=0.51$ for the \agn{} sample to
$Z_{\rm Fe,IN}=1.07$ for the \csf{}, as well as an extension of the
range covered especially towards high-metallicity values.
Additionally, we find an overall shift of the majority of the clusters
towards higher values of the pseudo-entropy $\sigma$.  Median values
of the distributions shown in Fig.~\ref{fig:entr-met0} are reported in
Table~\ref{tab:sigma}.  In the \csf{} run, the lack of an efficient
form of feedback, such as from AGN sources, prevents the formation of
realistic CC clusters and the majority of the clusters could be
classified as NCC systems\footnote{We note that the threshold
  $\sigma=0.55$ used to discriminate between CC and NCC was rather
  applied to 3D values of $\sigma$ in~\cite{rasia2015}. Despite
  different specific values of $\sigma$, we find, nonetheless, the
  same classification at $z=0$ when the projected values are used as
  in Fig.~\ref{fig:entr-met0}\,(top).}.
Nonetheless, such systems do not resemble realistic NCC clusters
  either, since they present both high entropy and high metallicity in
  the core, whereas observed NCC systems are rather characterized by
  low central metallicity.
Concerning the five clusters lying below the
threshold used to identify CC systems, we further investigated their
entropy maps and profiles and found that they showed substantial
differences with respect to the typical properties of realistic CC
clusters in the \agn{} run.  For a more detailed discussion, we refer
the reader to the Appendix.

\begin{table*}
\centering
\begin{tabular}{lllcccccc}
\hline
& && \multicolumn{2}{|c|}{2D [EW]} && \multicolumn{3}{|c|}{3D [MW]}\\
\hline
& && $\sigma$ & $Z_{\rm Fe,IN}/Z_{\rm Fe,\odot}$ && $\sigma$ & $Z_{\rm Fe,IN}/Z_{\rm Fe,\odot}$ & $Z_{\rm Fe,OUT}/Z_{\rm Fe,\odot}$ \\
\hline
\multirow{3}{*}{AGN}
& all && 0.68 & 0.51 && 0.81 & 0.47 & 0.28\\
& CC  && 0.41 & 0.58 && 0.42 & 0.50 & 0.28 \\
& NCC && 0.82 & 0.47 && 1.06 & 0.43 & 0.28\\
\hline
\multirow{1}{*}{CSF}
& all && 0.70 & 1.07 && 0.86 & 0.99 & 0.37\\
\hline
\end{tabular}
\caption{Median values of pseudo-entropy and metallicity for the
  \agn{} and \csf{} simulated samples at $z=0$. The distributions used
  for the projected values (columns 3 and 4) are the emission-weighted
  (EW) estimates shown in Fig.~\ref{fig:entr-met0}. For the
  three-dimensional case, the values reported are mass-weighted (MW)
  instead, and also the results for the ``OUT'' region are reported
  (columns 5, 6 and 7), as in the top panel of
  Fig.~\ref{fig:entr-met-evol}.}\label{tab:sigma}
\end{table*}

 \subsection{ICM metallicity profiles}\label{sec:met-profs}
In order to asses the reliability of our \agn{} simulations with respect to
the chemical properties of galaxy clusters observed in the local
Universe, we also investigate the radial distribution of iron and
oxygen-to-iron and silicon-to-iron abundance ratios.
 \begin{figure*}
 \centering
 \includegraphics[width=0.45\textwidth,trim=30 0 0 25,clip]{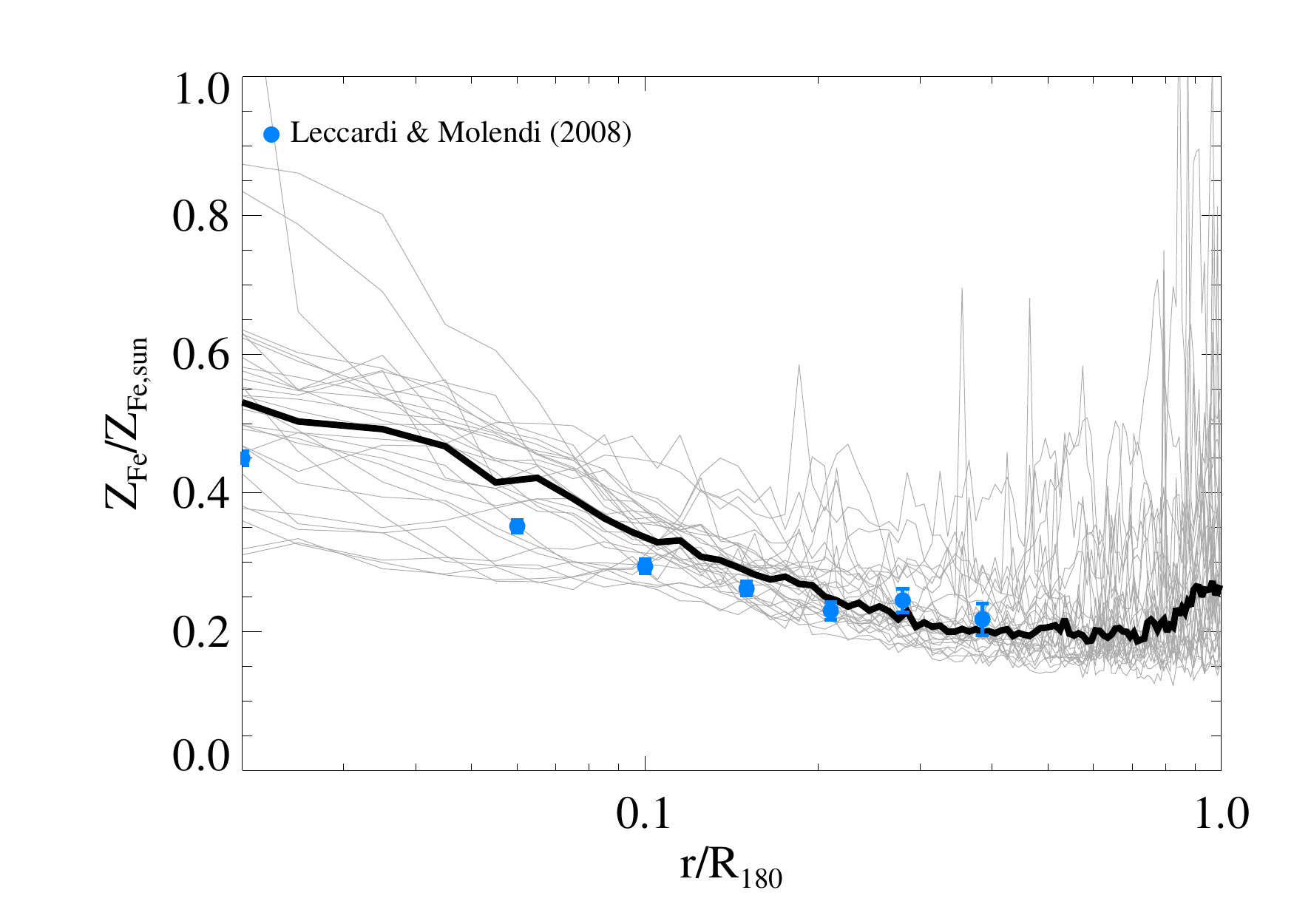}
 \includegraphics[width=0.45\textwidth,trim=30 0 0 25,clip]{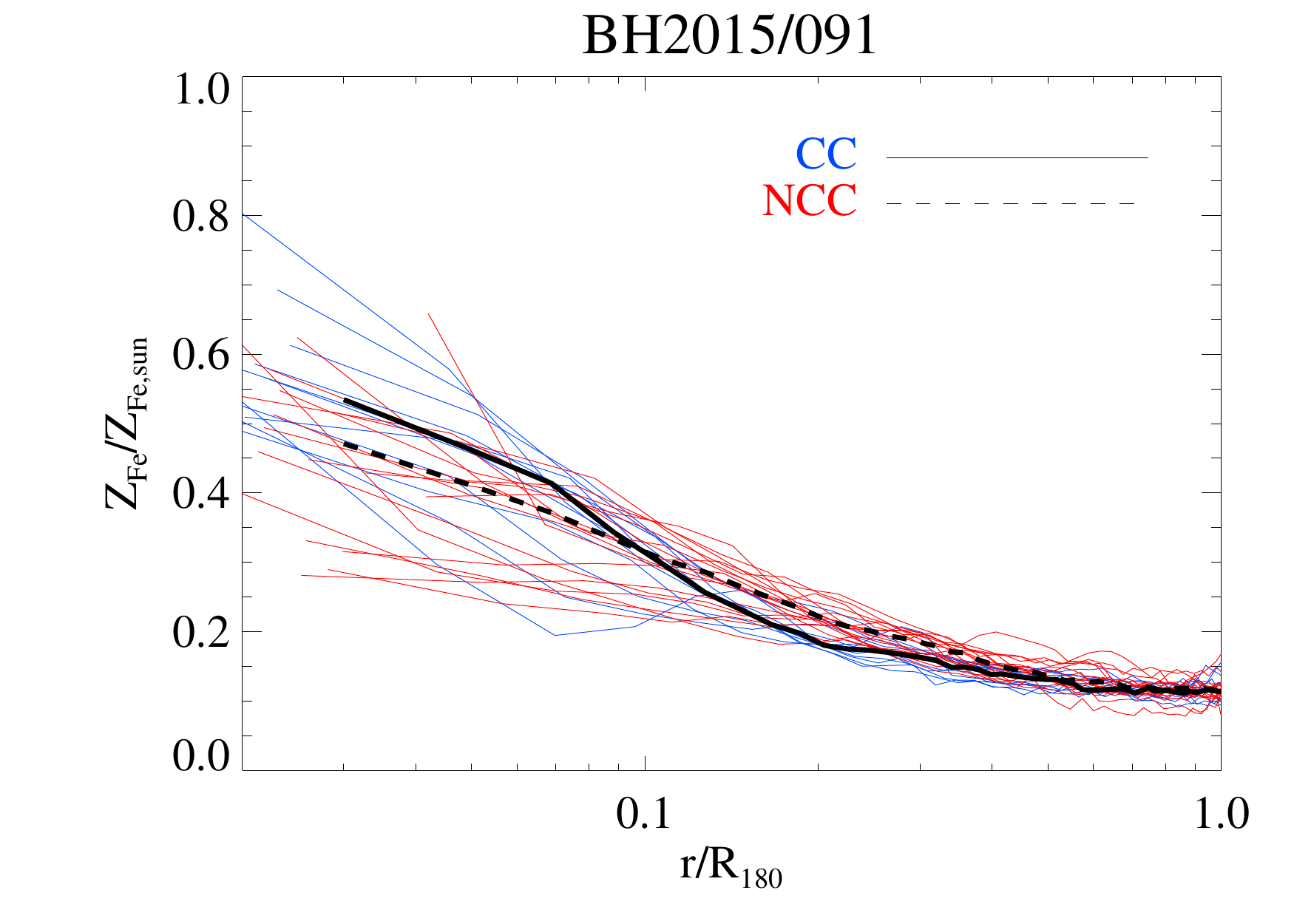}\\
 \caption{Radial profiles out to $R_{180}$ of the iron abundance, relative to
  solar, for our samples of clusters at $z=0$.
  {\it Left:} emission-weighted projected profiles compared against
  observational data by~\protect\cite{Leccardi_2008} (blue circles). Marked
 with the solid black line we show the median profile for the 29
  clusters. The integration along the l.o.s.\ covers a length of $2\,\rvir$.
{\it Right:} mass-weighted three-dimensional profiles for
 the CC (blue) and NCC (red) clusters separately, with the thick
  solid and dashed lines marking the median profiles of each
  subsample.}
\label{fig:profs0}
\end{figure*}

In the l.h.s. panel of Fig.~\ref{fig:profs0} we show in particular the
projected emission-weighted profile of iron abundance, relative to the
solar value, for the ICM. The black solid line refers to the median
profile across the sample.  Here, simulation predictions are compared
to observational data points by~\cite{Leccardi_2008} (blue data
points).  We find good agreement between simulated and observed
profiles, within the dispersion.
Towards the centre, metallicity values in the simulations appear
slightly higher and show a large scatter, with respect to
observational profiles. Nevertheless, we note that the results are
compatible, as the scatter around the observational data reported by
\cite{Leccardi_2008} is also high (see dashed lines reported on their
Fig.~2).  Large scatter in the innermost regions is also found in
recent high-resolution Chandra observations of single cluster profiles
shaped by AGN feedback~\cite[e.g.,][]{Kirkpatrick_2015}.

In the r.h.s. panel of the Figure, we present instead the
mass-weighted, three-dimensional metallicity profiles, which quantify
the intrinsic distribution of the iron-rich ICM.
Colors refer to the cool-coreness
classification~\cite[see][]{rasia2015} of the clusters, with CC
systems marked in blue and NCC ones in red. The thick solid and dashed
lines represent the median profiles of the two subsamples (CC and NCC,
respectively).
The 3D iron profiles show a lower Fe metallicity in the outskirts
as well as a reduced scatter.
This similarity of the iron content within the sample implies a
homogeneous distribution of metals in the outskirts.
In contrast, the larger variance seen in the emission-weighted profiles
is mainly due to the presence of substructures and gas density clumps,
especially at large cluster-centric distances,
which more significantly affect the estimates weighted by the ICM emission
(as in observations).
On the other hand, the metal content of the central regions reflects
the diversity of the cluster core properties.
Our sample of CC clusters shows on average higher metal
abundances and steeper profiles than NCC systems in the inner cluster
region ($r<0.1R_{180}$), which is in agreement with observational
findings by~\cite{ettori2015}, as shown and discussed in~\cite{rasia2015}.

\enlargethispage*{\baselineskip}
\subsubsection*{Abundance ratios}

In addition to studying the single chemical species, computing the
relative abundance of heavy elements produced via different enrichment
channels is also particularly useful in order to infer information on
the chemical history of galaxy clusters.
\begin{figure*}
\centering
\includegraphics[width=0.45\textwidth,trim=30 0 0 25,clip]{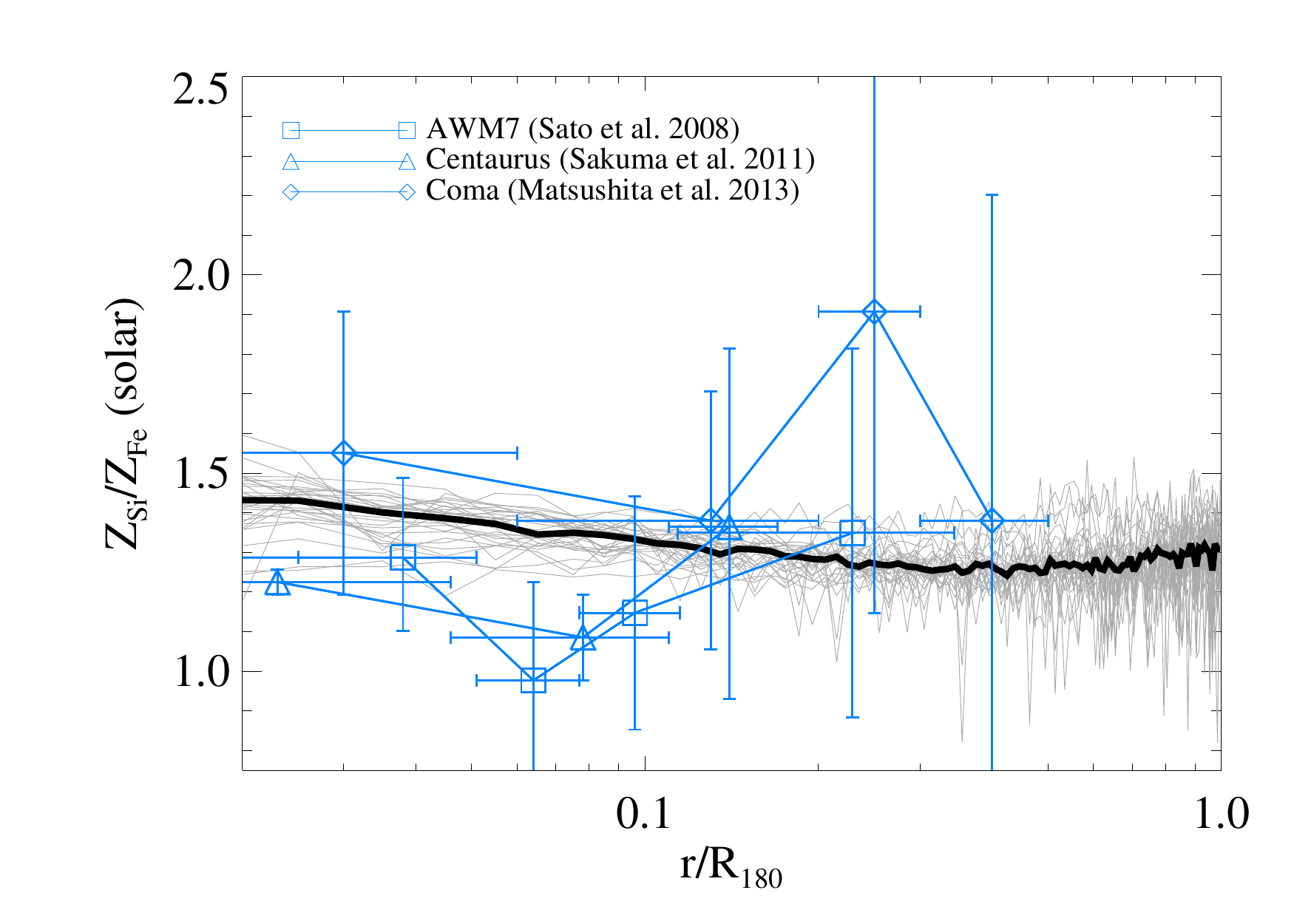}
\includegraphics[width=0.45\textwidth,trim=30 0 0 25,clip]{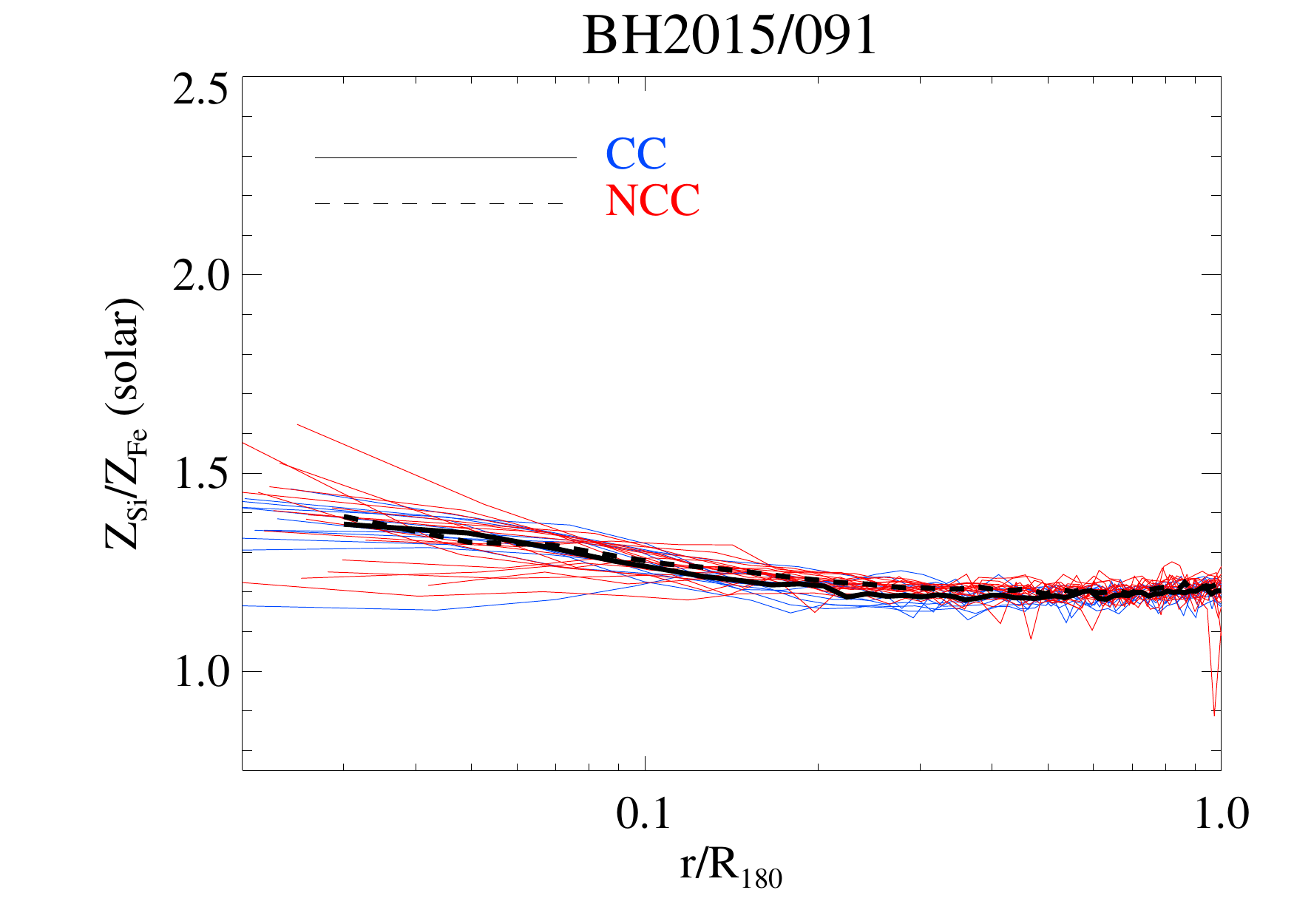}\\
\includegraphics[width=0.45\textwidth,trim=30 0 0 25,clip]{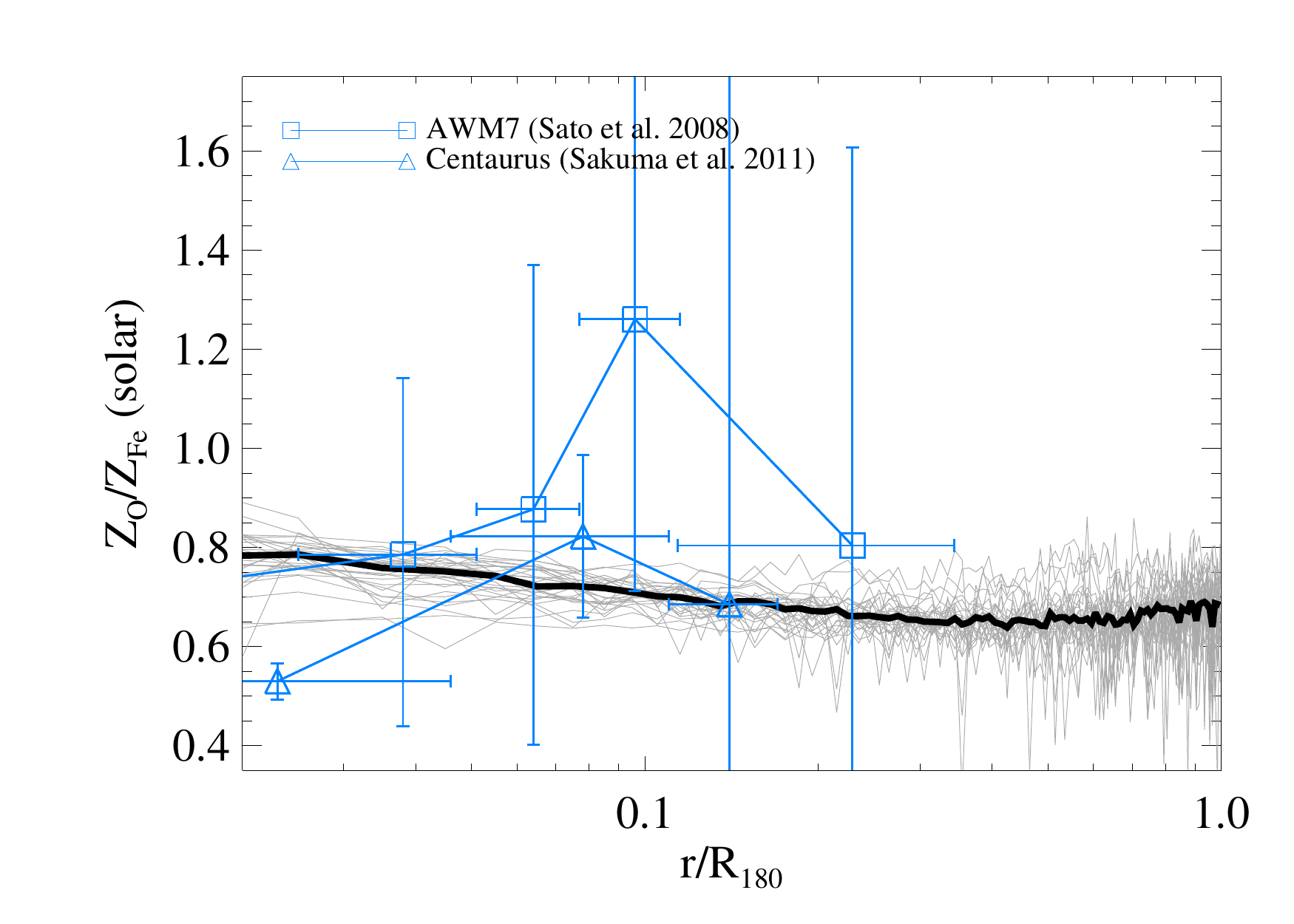}
\includegraphics[width=0.45\textwidth,trim=30 0 0 25,clip]{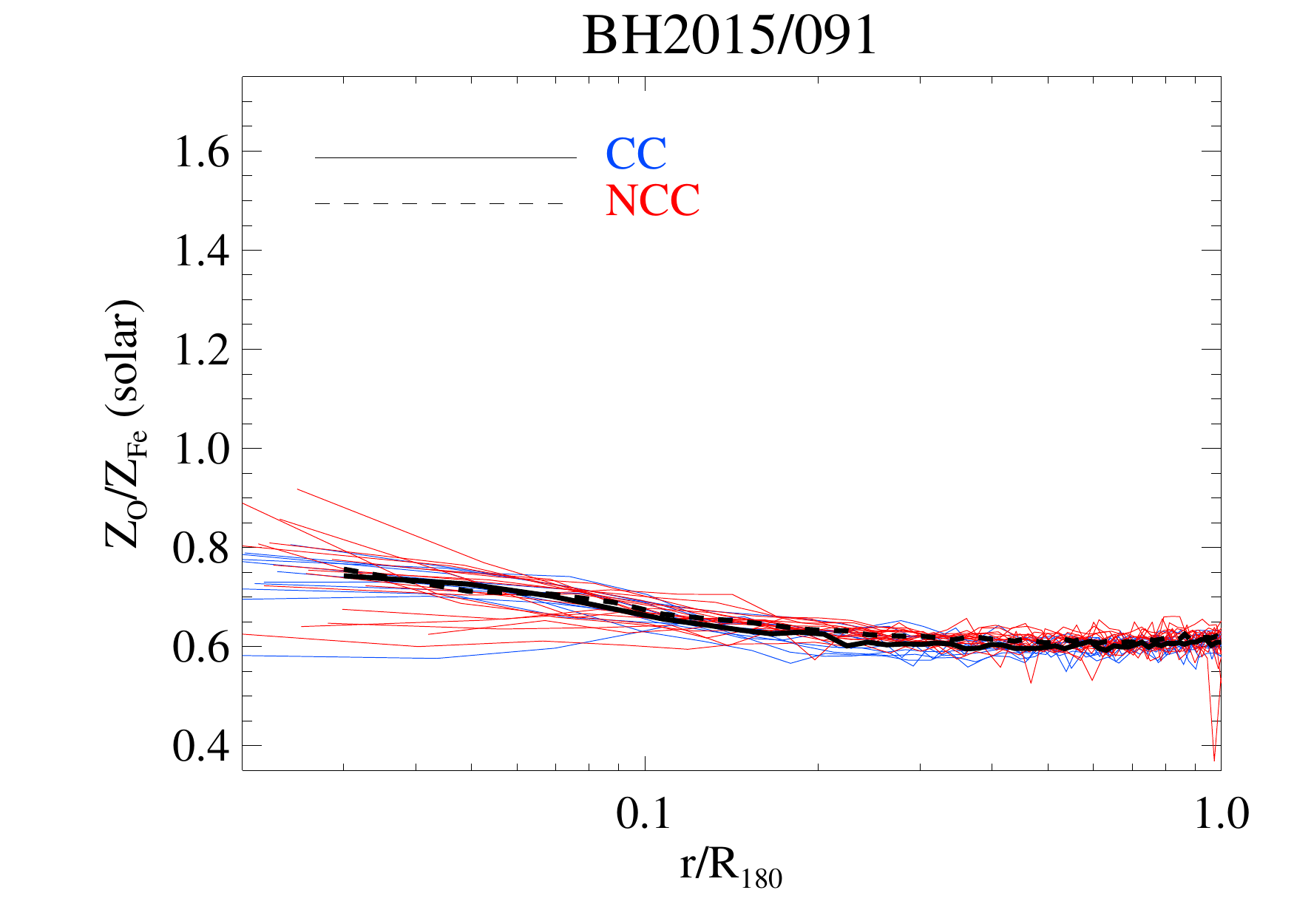}
\caption{Radial profiles out to $R_{180}$ of the relative
  abundances of Si/Fe (top panel) and O/Fe (bottom panel) for our
  samples of clusters at $z=0$. {\it Left:} projected
  emission-weighted profiles, and median profile (thick black curve),
  compared to observational data (blue symbols) from \suzaku\
  observations of three massive clusters, as in the legend. {\it
    Right:} three-dimensional mass-weighted profiles for the CC (blue)
  and NCC (red) clusters in the sample.}
\label{fig:profs0-2}
\end{figure*}

\looseness=-1
For our \agn{} sample, we show in
Fig.~\ref{fig:profs0-2} the radial profiles of the abundance ratio of
silicon and oxygen relative to Fe (Si/Fe in the top panels and O/Fe in
the bottom panels).
In the l.h.s. panels we report the projected, emission-weighted
profiles computed from the simulations and observational data from
\suzaku\ for the massive galaxy clusters: AWM7~\cite[][]{sato2008},
Centaurus~\cite[][]{sakuma2011} and --- only for the Si/Fe profile ---
Coma~\cite[][]{matsushita2013}.

With respect to observational data, we find an overall agreement, given
the large error bars on observed profiles.  In particular, our
findings confirm the flatness of the ICM metallicity profiles in the
outer cluster regions, indicating a homogeneous distribution of the
metal-enriched gas~\cite[see recent studies
  by][]{werner2013,simionescu2015}.

Interestingly, by inspecting the three-dimensional, mass-weighted
abundance ratio profiles in the r.h.s. panels of
Fig.~\ref{fig:profs0-2}, we confirm that the distribution of Si and O
relative to Fe is intrinsically flat outside of $\sim
0.1$--$0.2\,R_{180}$, with a very
small intrinsic scatter across the sample.

This picture suggests that most of the metal enrichment of the ICM,
from a combination of SNII and SNIa, occurred before the cluster
formed.
For this reason, it is crucial to explore the ICM enrichment history
in simulations, once their reliability in describing the observational
results at present time has been tested.

Notice that the intrinsic abundance ratio profiles of the CC and NCC
populations do not differ significantly, suggesting that the
enrichment due to SNIa and SNII has an overall similar distribution
in the two cases.
In fact, in our sample the behaviour of the silicon
and oxygen abundances is very similar to that of iron, shown in
Fig.~\ref{fig:profs0} (right): typically, the CC subsample shows steeper
radial profiles that are more peaked towards the cluster centre, while
the NCC systems present flatter profiles, which are higher than CC for
$0.1\,R_{180} < r < 0.3\,R_{180}$, and reach lower metallicities in the centre.
This is mirrored by the relatively flat profiles and by the lack of
distinctive signatures between CC and NCC systems of the Si/Fe and
O/Fe abundance ratios.


\section{The history of the ICM metal enrichment}\label{sec:evol-met}

Given the relatively good agreement with projected data at
$z=0$ (Fig.~\ref{fig:entr-met0}), we use the simulated clusters to
show how the relation between central entropy and metallicity evolves with time.
However, in order to make a prediction on the intrinsinc metal
enrichment of the ICM at higher redshifts, we take advantage of the
simulated data to show the mass-weighted metallicity values, and do
not project along the line of sight.

\begin{figure}
\centering
\includegraphics[width=0.45\textwidth,trim=30 7 20 5,clip]{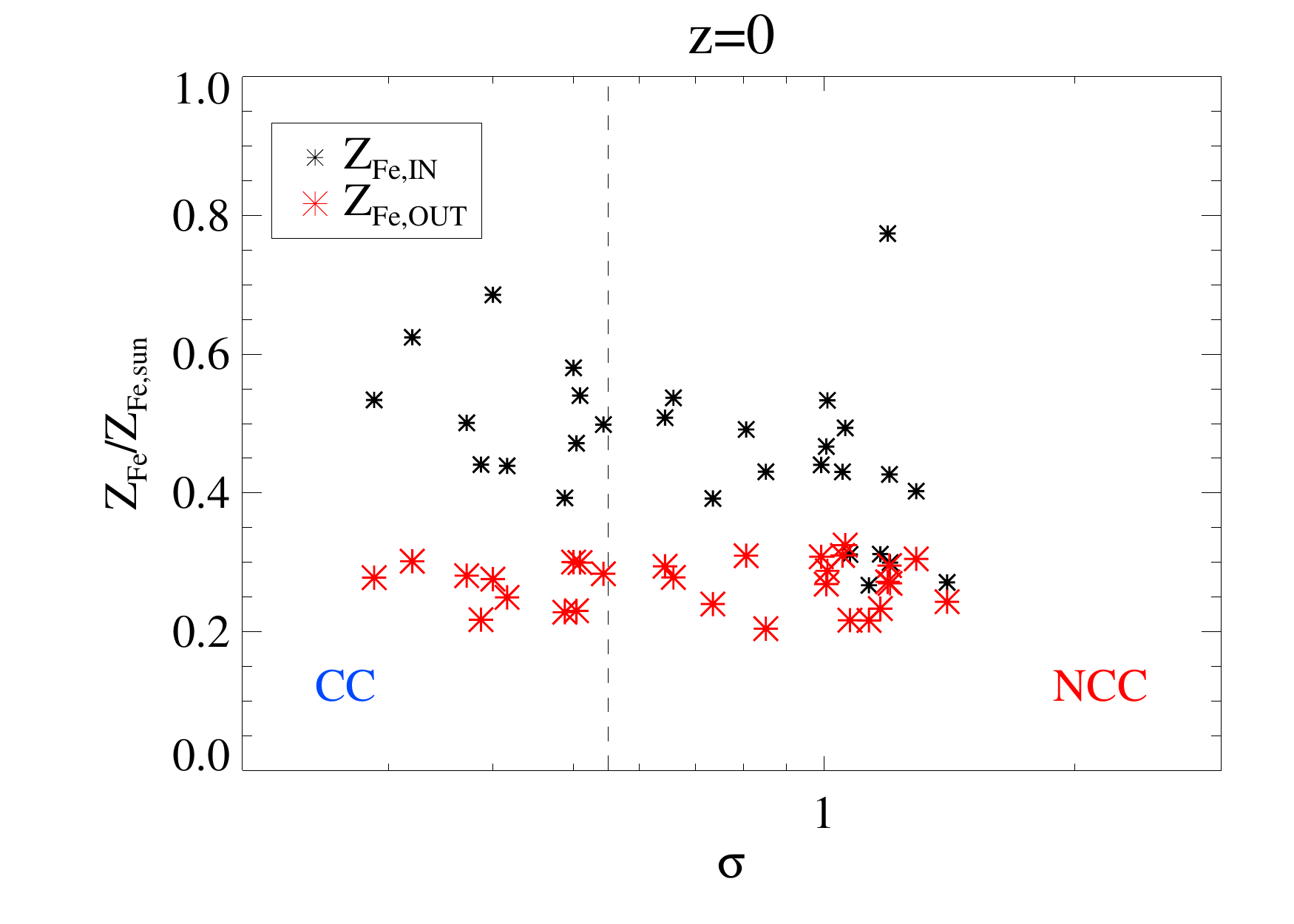}
\includegraphics[width=0.45\textwidth,trim=30 7 20 5,clip]{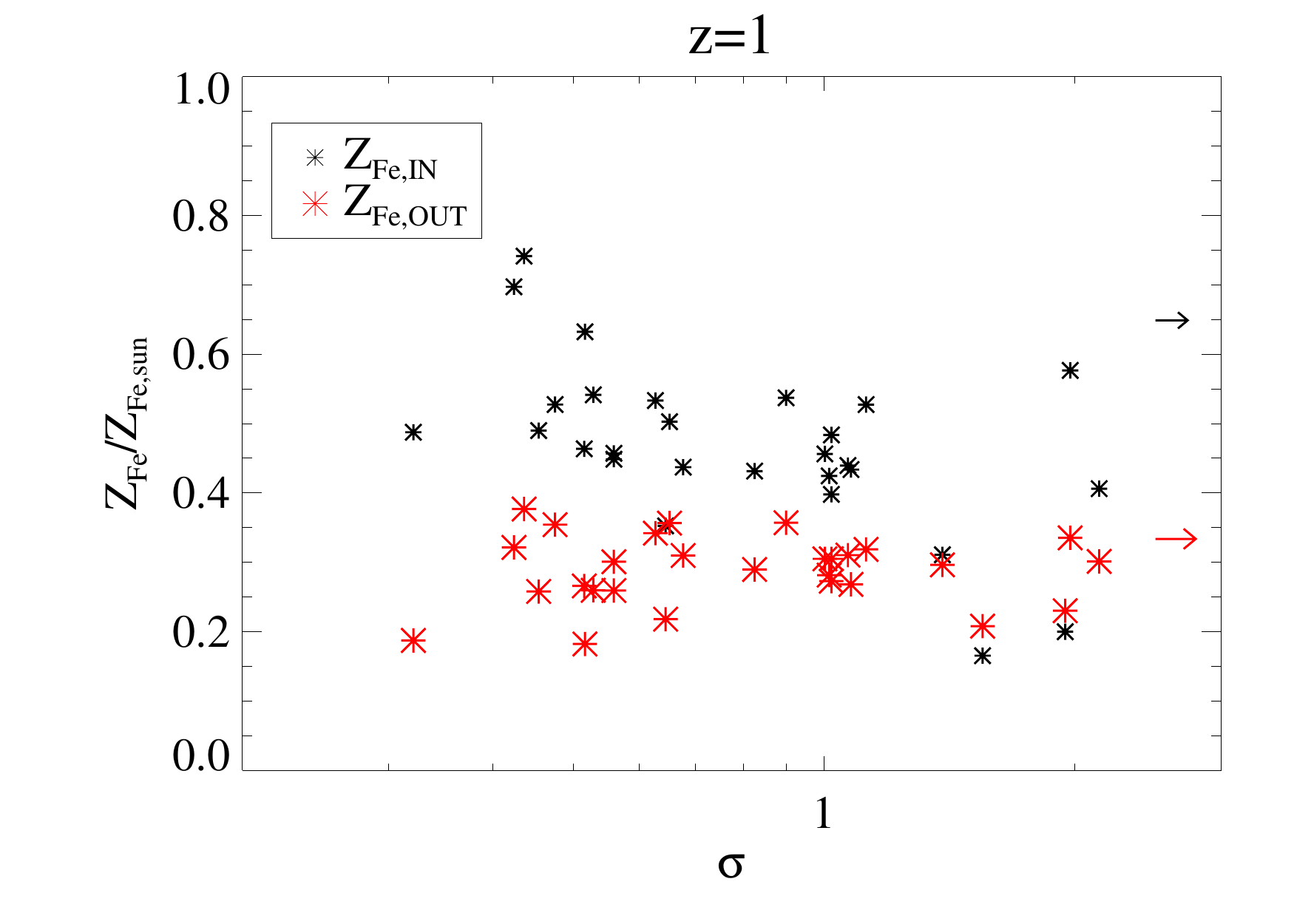}
\includegraphics[width=0.45\textwidth,trim=30 7 20 5,clip]{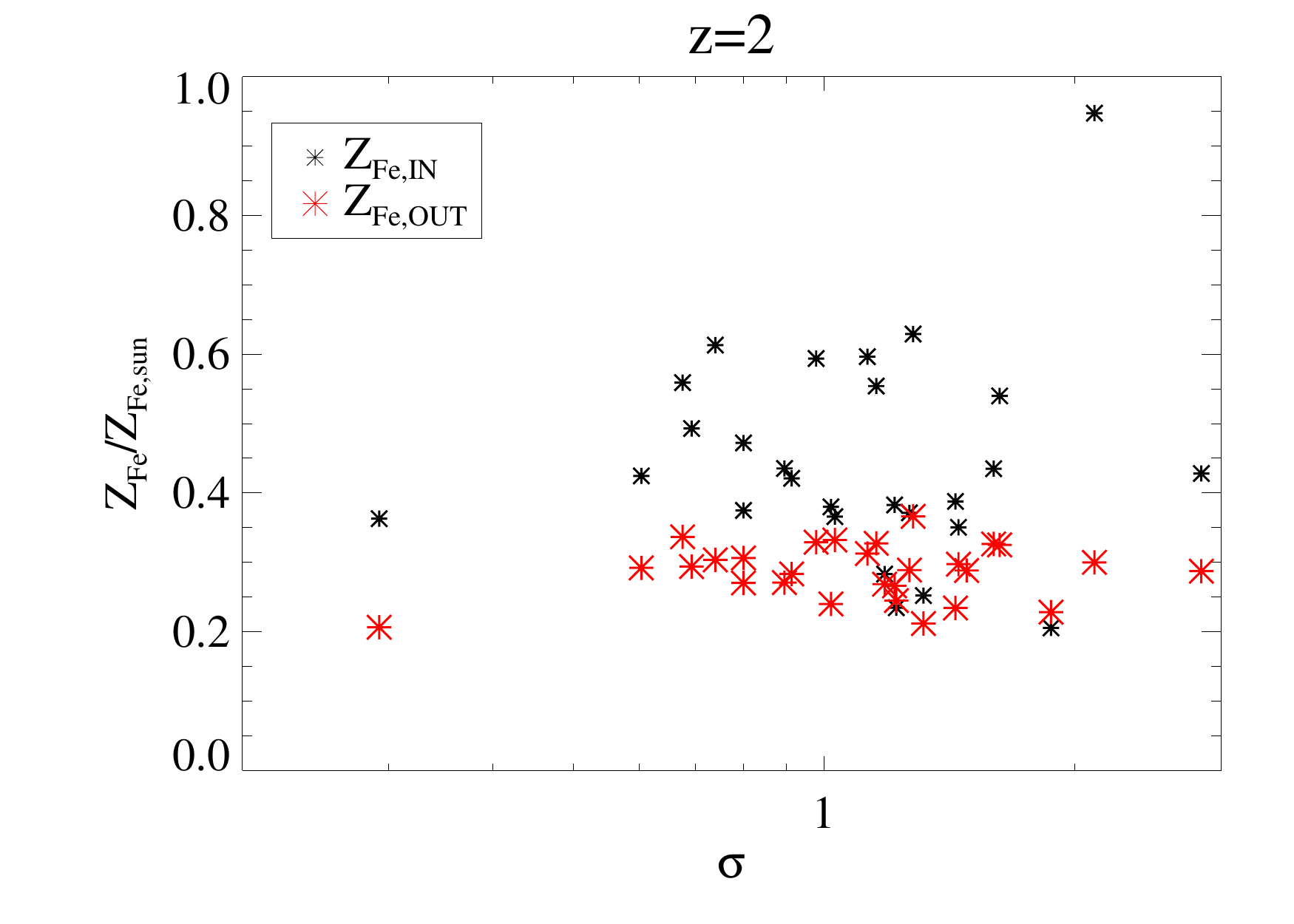}
\caption{Relation between mass-weighted metallicity and pseudo-entropy
  ratio for the simulated clusters at $z=0,1,2$ (from top to bottom).
  Values are computed within the three-dimensional ``IN'' (black asterisks)
  and ``OUT'' (red asterisks) regions, corresponding to
  $r<0.05\,R_{\rm 180}$ and $0.05\,R_{\rm 180} < r < 0.2\,R_{\rm
    180}$, respectively. At $z=0$ we report the
    $\sigma$-threshold used to discriminate between CC and NCC
    clusters (vertical dashed line).  The arrows in the central panel
    ($z=1$) indicate the metallicity values of a data point for which
    $\sigma$ lies outside the plotted range.}
\label{fig:entr-met-evol}
\end{figure}

In Fig.~\ref{fig:entr-met-evol} we show the evolution of the intrinsic
relation between mass-weighted metallicity of the innermost region of
our simulated clusters and the 3D pseudo-entropy ratio.
From the three panels in Fig.~\ref{fig:entr-met-evol},
we note that the intrinsic correlation between central metallicity
$Z_{\rm Fe,IN}$ and pseudo-entropy $\sigma$ (black symbols), albeit
still negative, is weaker than the one reported in Fig.~\ref{fig:entr-met0},
and essentially vanishes at $z=2$.  The mean value of the central
metallicity is only slightly increasing from $<Z_{\rm Fe,IN}>\sim
0.43$ at $z=2$ up to $<Z_{\rm Fe,IN}>\sim 0.47$ at $z=0$. In fact, we
note from the Figure that the main evolutionary effects are linked to
the change of the typical pseudo-entropy range with redshift, rather
than to any significant change of the intrinsic enrichment level of
the gas.
For comparison, we also show the evolution of the relation between the
pseudo-entropy $\sigma$ and the metallicity in the ``OUT'' spherical
shell, which is actually the intermediate region of the clusters
  (i.e.\ $0.05\,R_{\rm 180} < r < 0.2\,R_{\rm 180}$).
This is marked in the figure by the red symbols.
In this case we find an almost flat distribution around the median
value of $Z_{\rm Fe,OUT}\sim 0.28\pm 0.03\,Z_{\rm Fe,\odot}$ at
$z=0$. This trend is confirmed also at $z=1$ and $z=2$, with similarly
low scatter and mean values.  The intrinsic level of enrichment,
basically, is already established at $z\sim 1$--$2$,
especially in the ``OUT'' region, where the scatter
  across the different clusters of the sample is very~low.

\begin{figure}
\centering
\includegraphics[width=0.45\textwidth,trim=30 7 20 25,clip]{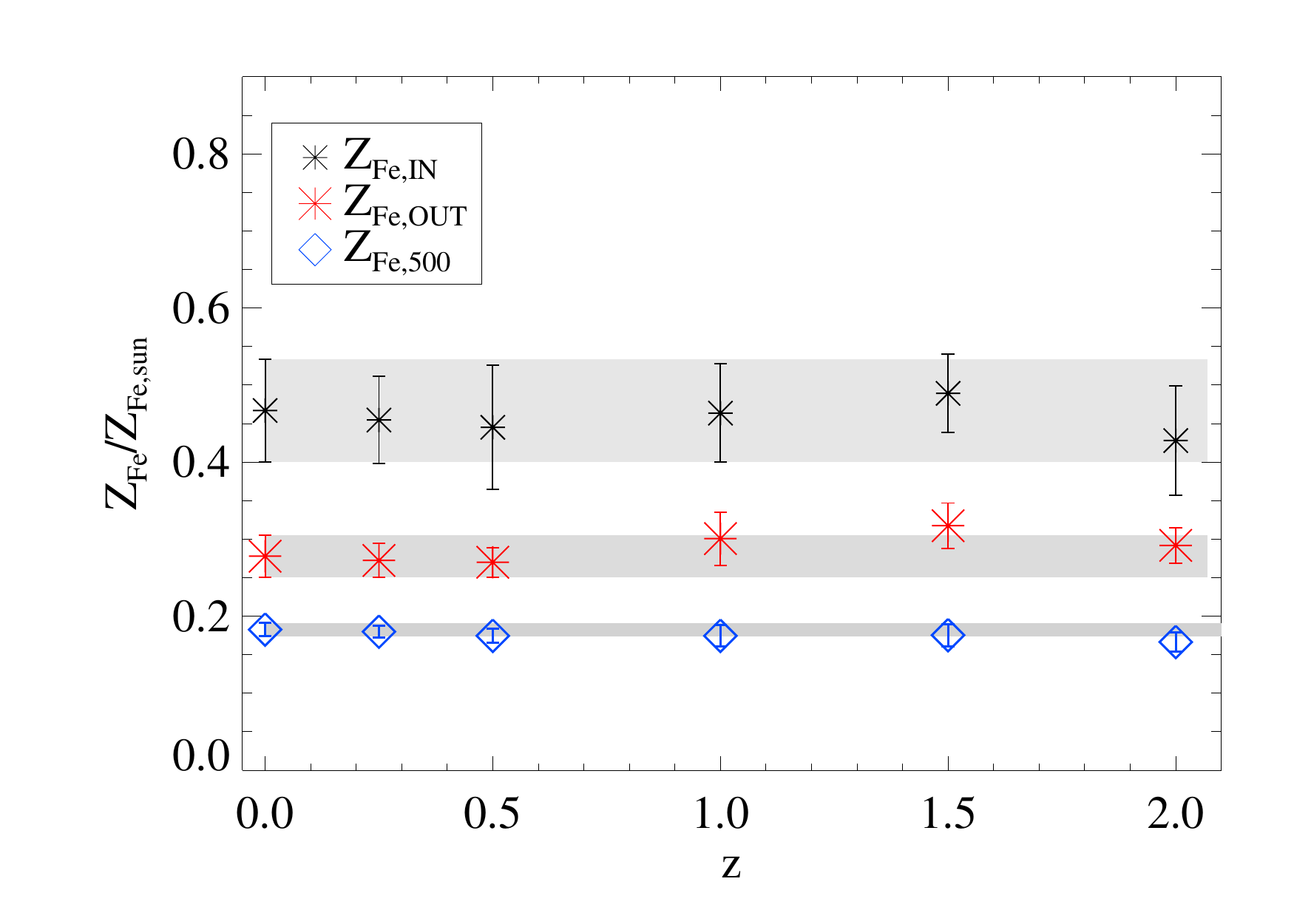}
\caption{Evolution with redshift of the median (intrinsic,
  mass-weighted) iron abundance within the ``IN'' (black asterisks)
  and ``OUT'' regions (red asterisks), corresponding to
  $R<0.05\,R_{\rm 180}$ and $0.05\,R_{\rm 180} < r < 0.2\,R_{\rm
    180}$, respectively. With blue diamonds we report the results for
  the $\rfive$ region.  The grey shaded areas mark the dispersion
  around the median values at $z=0$.}
\label{fig:entr-met-redshift}
\end{figure}

In Fig.~\ref{fig:entr-met-redshift} we show the evolution with
redshift, from $z=2$ to $z=0$ of the median Fe abundance in the two
considered regions, using the same color scheme.
The shaded areas marked in grey correspond to the dispersion around
the median value at redshift $z=0$ and show that there is very little
variation with redshift from what is observed at the present time.
Only at $z=1.5$ we note an increase in the typical metallicity, both
in the core and in the outer region, which is however still consistent with
the $z=0$ value, within the scatter.

Recent observational results by~\cite{ettori2015} showed a similar
picture on the spatial distribution of metals in the inner and outer
regions\footnote{In the analysis by~\cite{ettori2015}, the authors use
  slightly different definitions of the radial bins used to define the
  inner, intermediate and outer regions: $[0$--$0.15]\,\rfive$,
  $[0.15$--$0.4]\,\rfive$ and $>0.4\,\rfive$, respectively.  We
  verified that this difference does not affect our conclusions.} of
observed CC and NCC clusters in the redshift range $0.09 < z < 1.39$.
In particular, they also observe a neat decrease of the metal
abundance (by a factor up to $\sim 2$--$3$), moving from the inner
radial bin ($<0.15\,\rfive$) to the outer cluster regions
($>0.4\,\rfive$), especially for CC clusters. Instead, the
intermediate/outer cluster regions of both CC and NCC systems present
very similar average metallicity, and low intrinsic scatter.
In~\cite{ettori2015}, the authors also show observational evidences of
a correlation with redshift, indicating a negative evolution of the
central metal abundance with $z$, which is not observed in the
intrinsic metal enrichment of our simulated sample.  Our results are
also in overall good agreement with the findings
by~\cite{McDonald2016}, where the authors investigate the evolution of
the ICM metallicity in a sample of 153 galaxy clusters in the redshift
range $0 < z < 1.5$, observed with the \chandra, \xmm\ and
\suzaku\ telescopes.  In these observations, only the CC subsample
indicates a weak evolution of the central metallicity with redshift,
although the correlation is found to be less strong than
in~\cite{ettori2015}.  Instead, no significant difference is found
between CC and NCC systems when the core is excised and the metal
abundance is measured in the outer regions only, where no evolution is
found in both cases.  Also, the global metal abundance measured within
$\rfive$ does not show any strong evidence of evolution with redshift.
This is consistent with the trend observed in our simulated sample,
when the global (mass-weighted and three-dimensional) iron abundance
within $\rfive$ is concerned, as shown in
Fig.~\ref{fig:entr-met-redshift} (blue diamonds).

We note that in both observational works, the evolution in
the central metal abundance with redshift is mainly driven by the
evolution observed in the core metallicity of CC clusters.  In the
case of our simulated sample, we are not able to perform the analysis on the
separate CC and NCC subsamples, given the low statistics of CC at
higher redshift. Despite this difference, the common picture outlined
by both observations and simulations about the metal distribution in the
intermediate-outer regions is consistent with an
early-enrichment scenario, in which the bulk of the ICM metal content
was produced by a mixture of SNII and SNIa, widespread and incorporated at early
times ($z > 2$), even earlier than the peak of star-formation and AGN
activity. This is also consistent with the flat radial metallicity ---
and abundance ratio --- profiles at large distances from the cluster
centre.

\subsection{Evolution of the ICM metallicity profiles}

\begin{figure}
\centering
\includegraphics[width=0.44\textwidth,trim=50 15 20 35,clip]{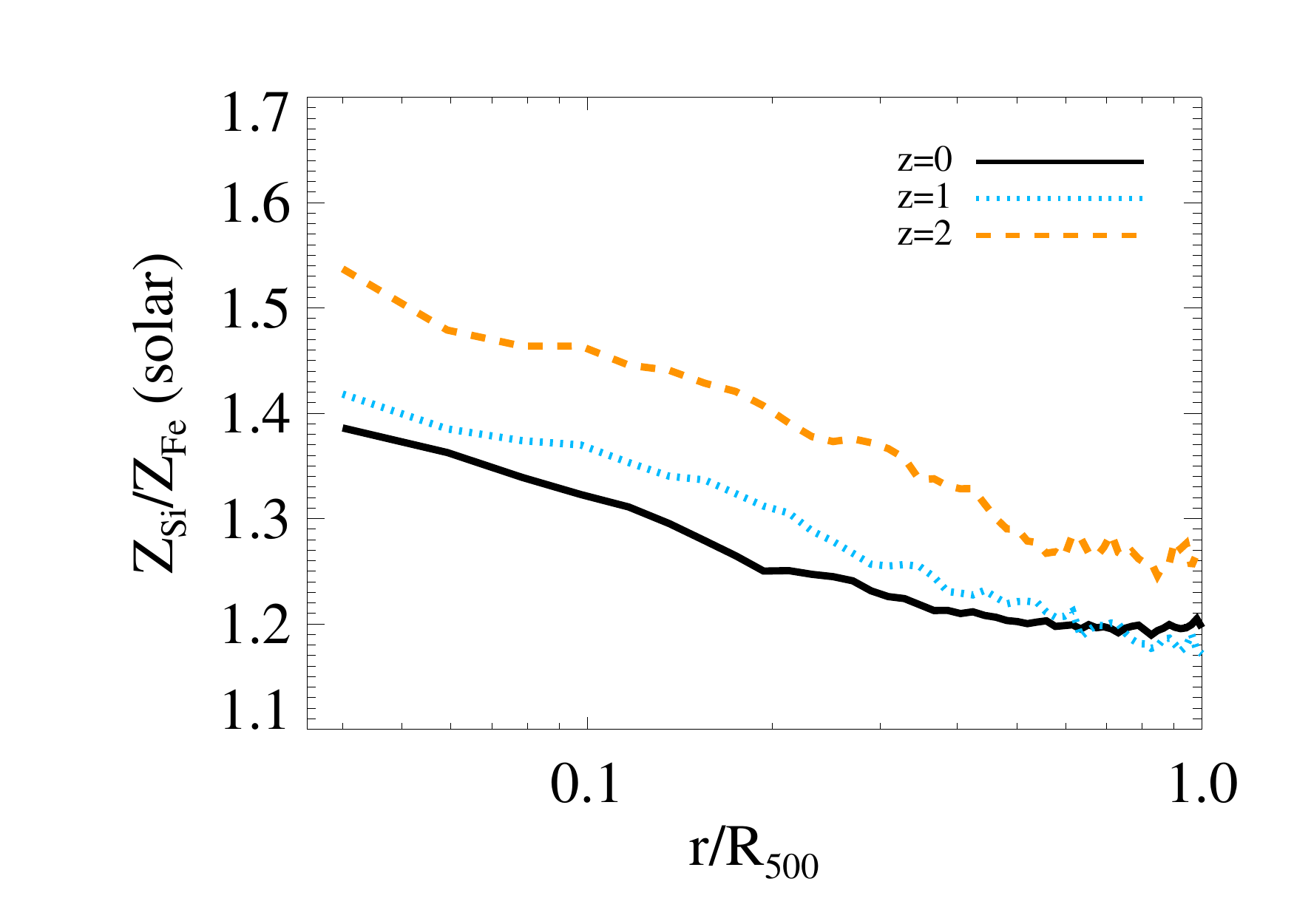}\\
\includegraphics[width=0.44\textwidth,trim=50 15 20 35,clip]{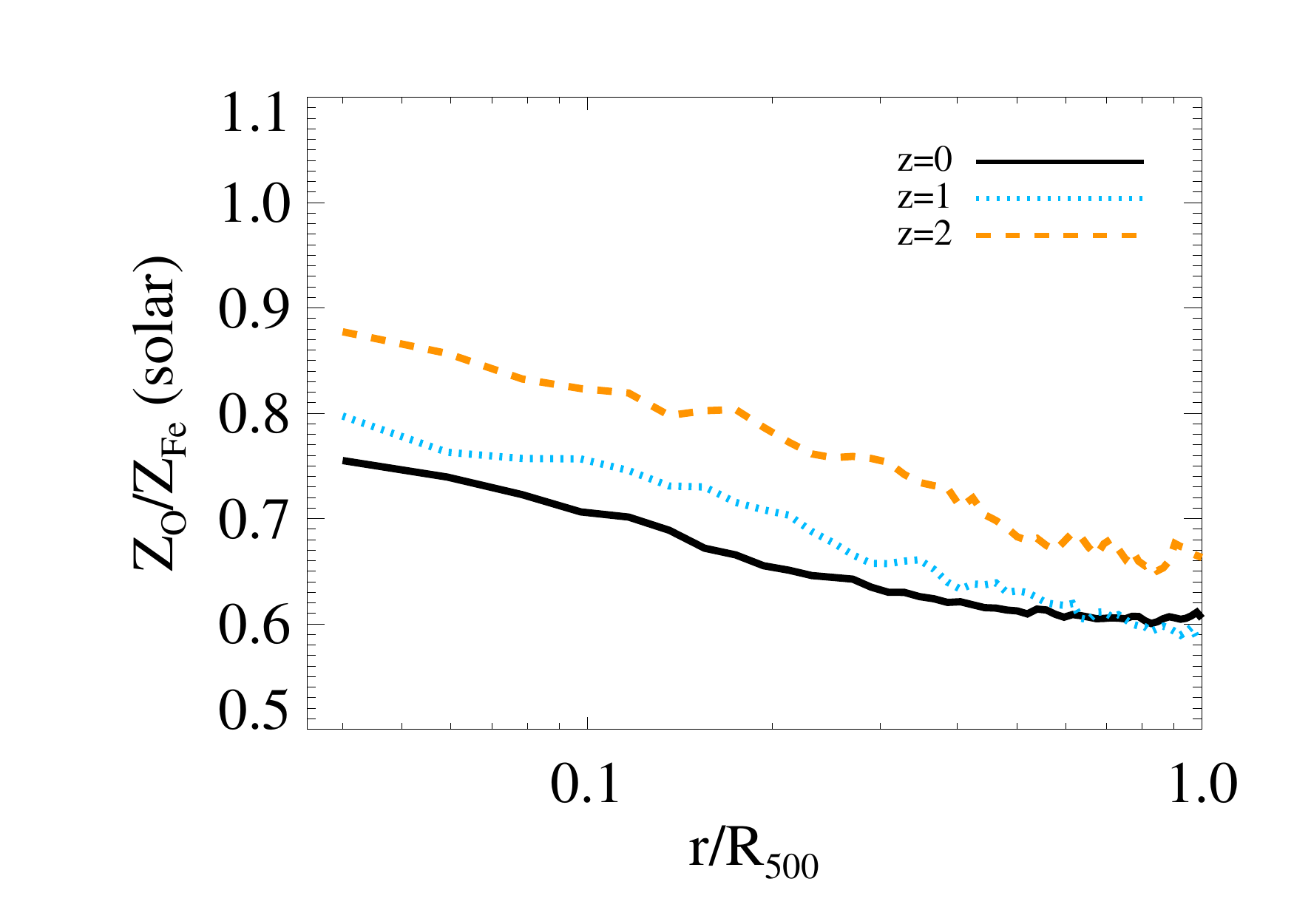}
\caption{Evolution of the median Si/Fe (top panel) and O/Fe (bottom
  panel) profiles from redshift $z=2$ to $z=0$. The profiles are
  three-dimensional and the quantities mass-weighted.
\label{fig:profs-evol}}
\includegraphics[width=0.47\textwidth,trim=50 10 20 20,clip]{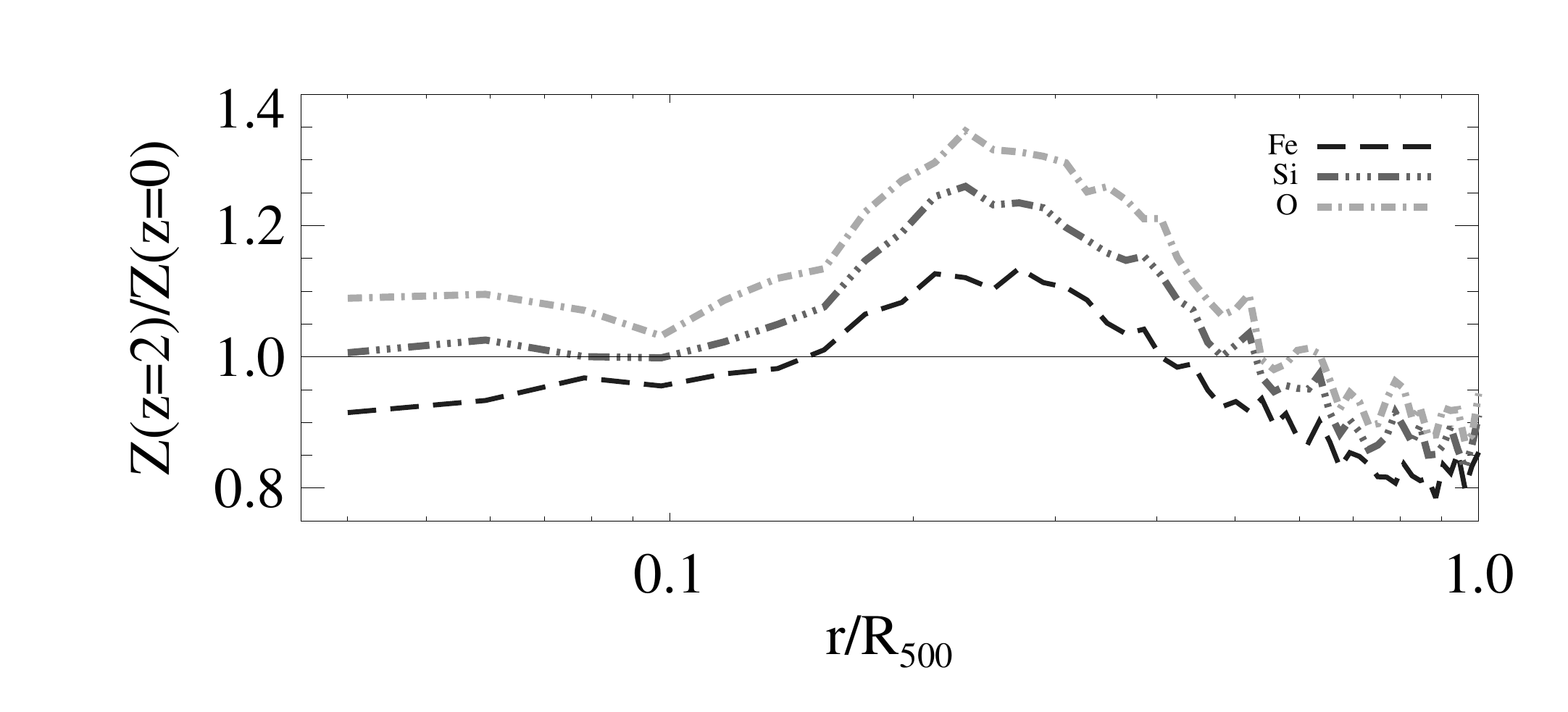}
\caption{Comparison between the median radial profiles at $z=2$ and $z=0$, for
  Fe, Si and O, separately. As above, the profiles are
  three-dimensional and the quantities mass-weighted.
\label{fig:fe-si-o}}
\end{figure}

In Fig.~\ref{fig:profs-evol}, we show the evolution of the radial
profiles of the Si/Fe and O/Fe abundance ratios, from $z=2$ to $z=0$.
In order to make predictions on the intrinsic evolution of the metal
enrichment of the ICM, we consider the behaviour of the median
profiles computed from the whole sample of 29 clusters, where
quantities are mass-weighted and profiles are three-dimensional.

\looseness=-1 From the Figure, we note that the ratio of abundance
profiles have similar slopes at all redshifts, although there is an
obvious evolution in the normalization. The most significant
difference is at $z=2$, when the abundance ratios are typically higher
than at lower redshifts, whereas between $z=1$ and $z=0$ we note a
weaker evolution.  With respect to $z=0$, the $z=2$ O/Fe abundance
ratio is on average $\sim 19\%$ and $\sim 10\%$ higher, at $r\sim
0.15\rfive$ and $r\sim \rfive$ respectively. For the Si/Fe abundance
ratio, the difference is $\sim 12\%$ and $\sim 6\%$ at the same
cluster-centric distances.

This general picture indicates that the abundance of iron relative to
elements produced by SNII, such as oxygen and silicon, increases from
$z=2$ to $z=0$. The reason for this trend is the longer timescale that
characterizes the enrichment from SNIa, associated to long-lived
stars, whose contribution (and iron production) becomes therefore more
and more important as time goes by.

In order to make this evolution clearer, in Fig.~\ref{fig:fe-si-o} we
also report the evolution of the profiles of the three single species,
singularly. From this, we note a clear difference in the normalization
of the $Z(z=2)/Z(z=0)$ ratio, for Si, O and Fe.
In the core region, the abundance of the individual
species does not vary significantly from $z=2$ to $z=0$, in fact there
is almost no evolution for the Si profile, while oxygen and iron show
a $10\%$ variation at most.  Nevertheless, the $z=2$ profile of Si and
O is similar or higher than the profile at $z=0$, while the iron
abundance in the same region (out to $\sim 0.15\rfive$) is instead
lower at $z=2$ than at $z=0$.  This opposite trend makes the
difference between the O/Fe (Si/Fe) abundance ratio at $z=2$ and that
at $z=0$ up to $20\%$ ($10\%$) in the innermost region.  Overall, the
systematic off-set between the Si and O curves with respect to Fe,
shown in Fig.~\ref{fig:fe-si-o}, is at the origin of the 10--20\%
evolution of the O/Fe and Si/Fe abundance ratios between redshifts 2
and 0.  The curves in Fig.~\ref{fig:fe-si-o} also denote that there is
a general tendency for the profiles to become from steeper at $z=2$ to
flatter at $z=0$.

\subsection{On the evolution of the baryonic total metal content}

Combining the results from the previous sections, the metallicity of
the hot-phase ICM at the centre (within the ``IN'' region,
i.e. $<0.05\,R_{\rm 180}$) shows a very mild increase of the central
Fe abundance (about $10\%$ from $z=2$ to $z=0$), that is however
consistent with no evolution, given the scatter across the sample.
The profiles shown in Fig.~\ref{fig:fe-si-o} are consistent with this
picture, and with the evolution of the absolute metal content of gas
and stellar component in the simulations.

\begin{figure}
\centering
\includegraphics[width=.43\textwidth,trim=35 0 25 0,clip]{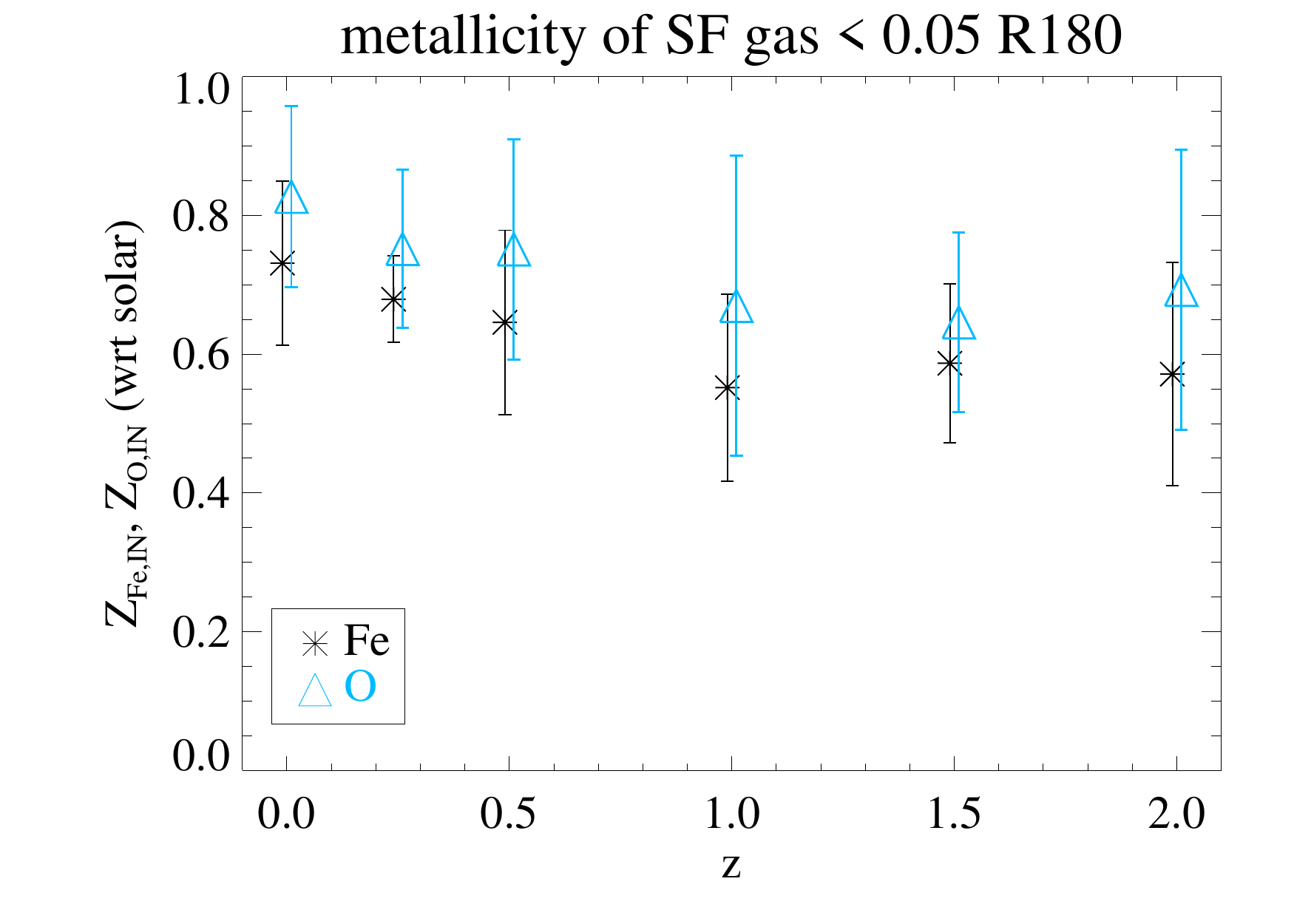}
\includegraphics[width=.48\textwidth,trim=5 0 10 0,clip]{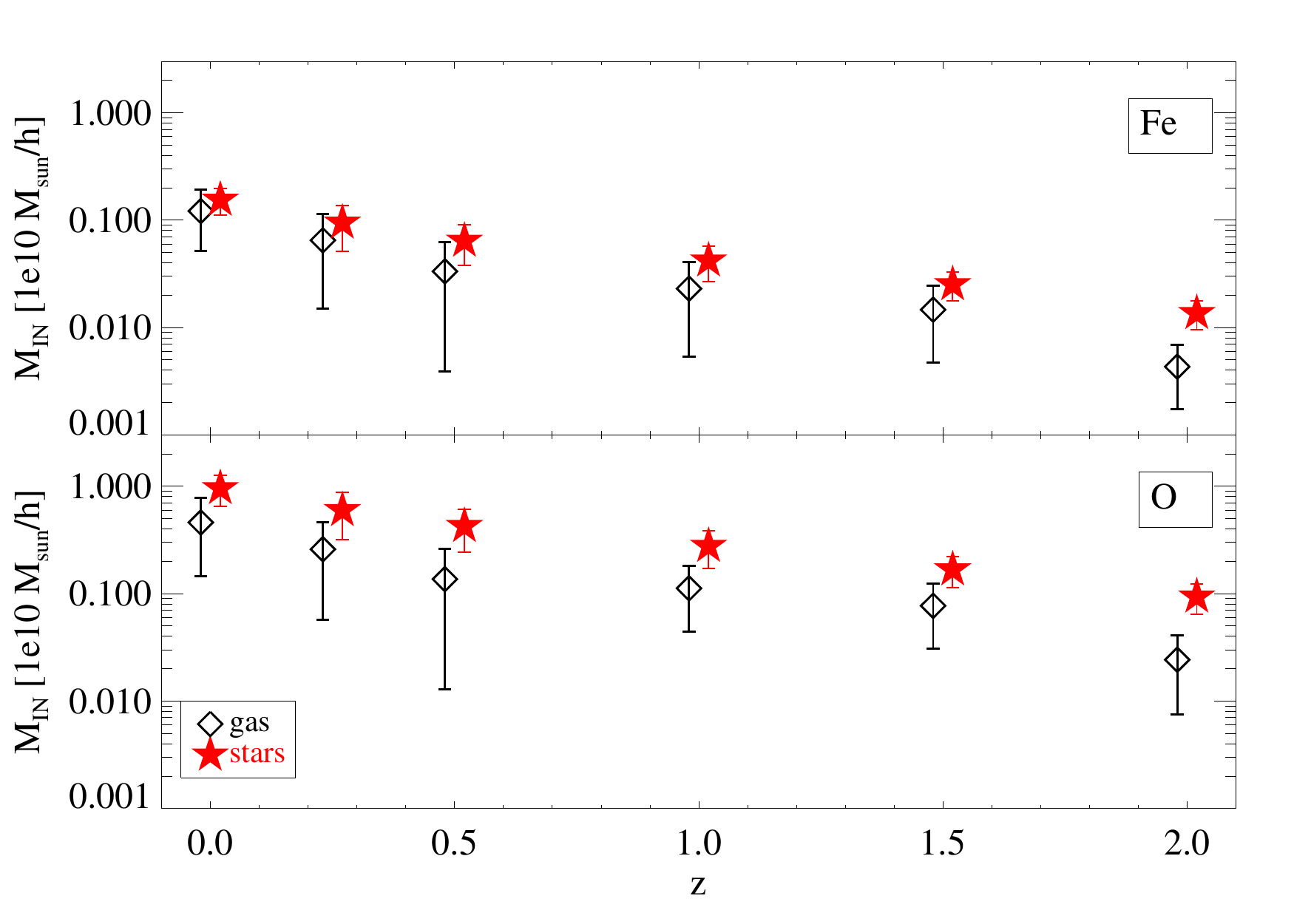}
\caption{\label{fig:metmass}{\it Upper panel:} evolution of the iron
  (black asterisks) and oxygen (cyan triangles)
  abundances~\protect\cite[wrt to Solar abundances by][]{angr1989}
  between $z=2$ and $z=0$, for the star-forming gas residing at each
  redshift within $0.05\,R_{\rm 180}$.  {\it Lower panel:} redshift
  evolution of the total iron (upper inset) and oxygen (lower inset)
  mass contained respectively in the gas (hot and SF; black diamonds) and stellar
  (red filled stars) components, residing within $0.05\,R_{\rm 180}$.
  In both panels, symbols indicate the median value and the scatter
  across the 29 clusters, at each redshift $z$.}
\end{figure}

In fact, this is due to the combined effect of lower-metallicity gas
collapsing to more central regions during the evolution of the
cluster, and of residual star formation consuming preferentially the
highly-enriched gas.  The former contributes to maintain or even
decrease the average metallicity, whereas the latter selectively
removes the gas enriched by either SNII or SNIa.  In the upper panel
of Fig.~\ref{fig:metmass}, we plot the evolution from $z=2$ to $z=0$
of the median (mass weighted and 3D) iron and oxygen abundances, for
the star-forming gas in the ``IN'' region.
According to the effective star-formation model implemented in our
code~\cite[][]{springel2003}, the star-forming gas is identified with
the gas particles whose density exceeds the star-formation density
threshold and are therefore treated as multi-phase (i.e. they contain
both a hot and a cold phase in pressure equilibrium, with the latter
being the reservoir for star formation ---
Sec.~\ref{sec:chem-mod}).\footnote{Essentially, the star-forming
  particles are those characterized by non-zero star formation rate,
  $SFR > 0$.}
Fig.~\ref{fig:metmass} (upper panel) shows that the
star-forming gas, from which new stars continue to form, is typically
highly enriched, especially with oxygen.  The difference between
oxygen and iron is still related to the different enrichment channels:
SNII produce a greater amount of oxygen than iron and given the short
time scale of the enrichment they likely pollute gas in the vicinity
of the star-forming region.  Such gas, dense and enriched, is
therefore also more subject to form new stars, before moving away due
to dynamical processes.  This effect is less prominent for Fe, since a
significant amount of Fe is produced by SNIa, whose longer time scale
allows the enrichment to happen also farther away from the
star-forming regions.

Despite these effects on metallicity evolution, the total mass of
metals in the gas component augments nevertheless with time, as
expected, due to the metal ejection accompanying the evolution of the
stellar component.  As a consequence of what discussed with the upper
panel of Fig.~\ref{fig:metmass}, also the mass of metals locked within
the stellar component increases with time, due to the formation of new
stars from the highly enriched gas.
In fact, considering the stars and the gas within the ``IN'' region,
the amount of metals in the stars even exceeds the one in the gas,
with a larger difference (by a factor of $\sim 2$ at $z=0$) in the
case of oxygen.  This is shown in the lower panel of Fig.~\ref{fig:metmass}
for the gas and the stars within $0.05\,R_{\rm 180}$ at each redshift,
for Fe (upper inset) and O (lower inset) separately.
Of the total mass of metals locked into the stars residing within the
central $0.05\,R_{\rm 180}$ at $z=0$, we verified that a
non-negligible fraction, about $\sim 35\%$ on average, belongs to
stars that formed at redshift $z<2$.

Therefore, we remark here that the low level of evolution found for
the median iron abundance of the ICM within the central ``IN'' region
strictly concerns the hot-phase gas component in the simulated
clusters, which is the one responsible for the X-ray ICM emission in
observed clusters.
Moreover, the lack of significant evolution of the central metallicity
for the whole sample is strictly related to the low statistics of CC
at higher redshifts, for which the evolution is observed in real
clusters.

This trend observed in the core region of clusters is confirmed on
larger scales. As shown in Fig.~\ref{fig:entr-met-redshift}, similar
investigations of the ``OUT'' and $\rfive$ regions show in fact no
evolution of the median metallicity of the hot gas with redshift. Also
at larger radii, the gas in the vicinity of star-forming regions will
be highly enriched with respect to the average, but will also be more
likely and efficiently converted into new stars. Similarly to what
happens in the core, the heavy elements in the metal-rich star-forming
gas will be locked into the newly formed stars without contributing to
increasing the metallicity of the hot gas.  Moreover, given the
accretion and infall of metal-poor gas, a certain balance is achieved
also on larger scales and the average metallicity is not expected to
increase significantly with time~\cite[see also][]{Martizzi_2016}.

\section{The effect of the AGN feedback}\label{sec:AGN-CSF}

\begin{figure*}
\centering
\includegraphics[width=.9\textwidth]{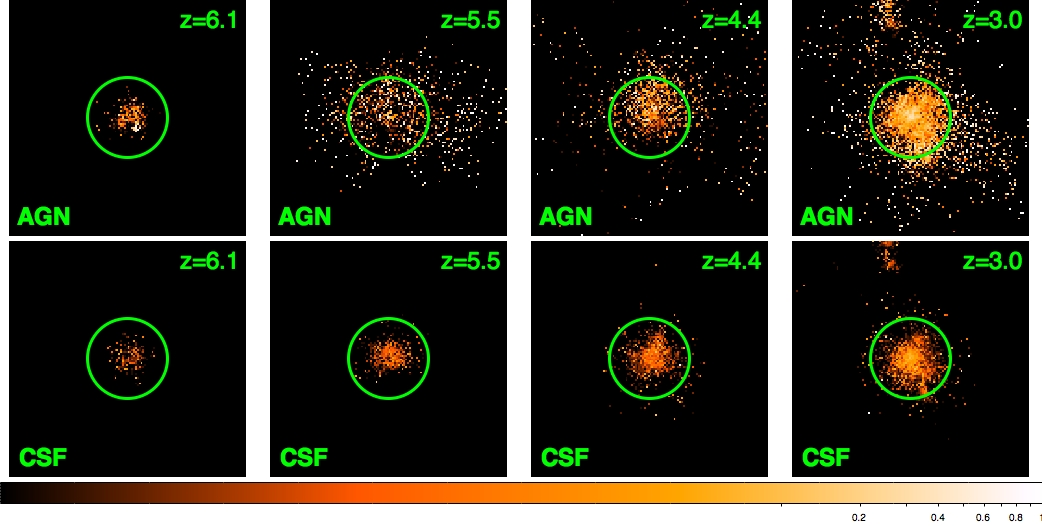}\\
\mbox{$Z_{O}/Z_{O,\odot}$}\\[10pt]
\caption{Oxygen maps at high redshifts for one example cluster in the
  runs with (top, \agn{}) and without (bottom, \csf{}) AGN
  feedback. From left to right, the redshifts considered are $z\sim
  6.1, 5.5, 4.4, 3$. The maps cover a region of $6\,\rvir$ per side,
  and $2$ virial radius along the projection axis, centred on the
  most massive halo at each redshift. The size of the region comprised
  within $\rvir$ is shown by the green circle.\label{fig:O-maps}}
\includegraphics[width=.9\textwidth]{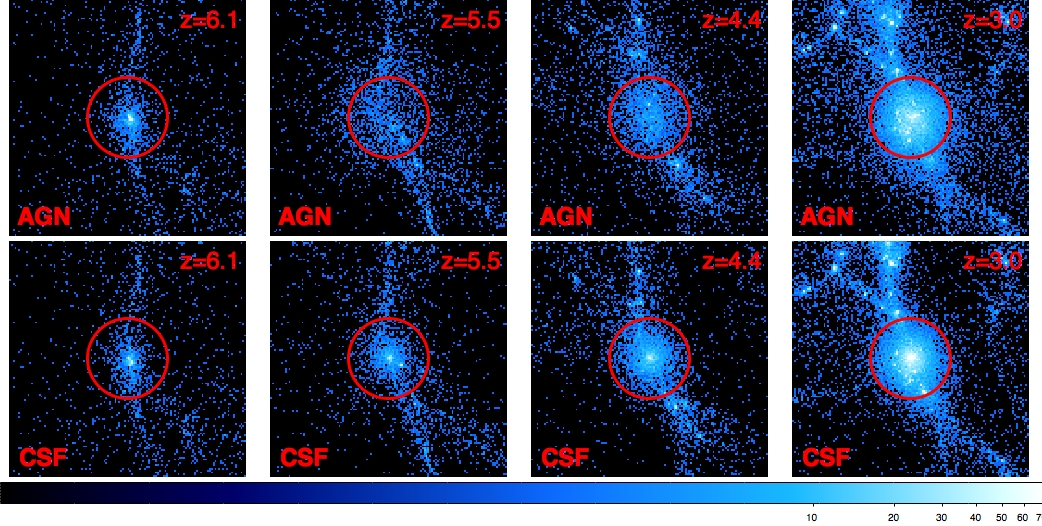}\\
\mbox{SFR [$M_{\odot}/yr$]}\\[10pt]
\caption{Similar to Fig.~\ref{fig:O-maps}, but for the total
  (projected) SFR in the same region. This basically traces the
  presence of star-forming regions.\label{fig:SFR-maps}}
\end{figure*}

The results discussed so far indicate a relatively homogeneous
distribution of metal-rich gas in the outskirts of low-redshift
clusters ($z\lesssim 2$), even concerning SNII products such as
oxygen.  Since the distribution of metal-rich gas is very
different in the case of \csf{} simulations, as visible from the
comparison shown in Fig.~\ref{fig:maps}, AGN feedback
has to be crucial in shaping the metal distribution observed at
the present time.

The origin of this different distribution in \agn{} and \csf{}
clusters can be better understood by inspecting the spatial
distribution of the metal-rich gas at higher redshifts.  The oxygen
abundance maps in Fig.~\ref{fig:O-maps} display the distribution of
oxygen around the most massive halo
 at $z\sim6$ in one of our Lagrangian regions, followed
 from $z\sim6$ down to $z\sim 3$,
for the two runs with and without AGN feedback (upper and lower
panels, respectively).  Each map encloses the region within $3\,\rvir$
from the halo centre
(circles mark the virial radius of the halo at each redshift), projecting the quantities along $2\,\rvir$ in
the l.o.s. direction.

Already at very high redshift, $z\sim 5.5$, the difference is clearly
visible: oxygen appears to be more widely spread when AGN feedback is
present.  The spatial spreading of oxygen-rich
gas happens in the \agn{} run during the passage from $z\sim 6$ to
$z\sim 5.5$, whereas no such spread is observed in the \csf{}
counterpart.  At $z\sim 5.5$, the
gas with $Z_{\rm O}\sim Z_{\rm O,\odot}$ has reached
distances beyond $ 2\,\rvir$ from the halo centre,
i.e.\ almost the map border,
in the \agn{} run.
On the contrary, the \csf{} cluster shows a much more concentrated
distribution of oxygen, whith the average metallicity increasing
mainly in the most dense and central regions, as the system evolves.

\looseness=-1 The metal distribution can be cross-correlated with that of
star-forming regions at the same redshifts, shown via the
star formation rate (SFR) maps in
Fig.~\ref{fig:SFR-maps}.  Focusing on the two left-most panels, we see
that in fact the SFR in the  innermost regions of the \agn{} system
is reduced while passing from $z\sim 6$ to $z\sim 5.5$, differently than in the \csf{} case.
At lower redshift, not only the SFR in the central regions of the \csf{}
halo is always higher than in the \agn{} case, but we also notice that the
distribution of oxygen continues to be typically confined to the
high-SFR regions, whereas it is spread over larger areas in the \agn{}
case. This is particularly evident at $z\sim 3$ (right-most panels),
where the SFR map is spatially very similar in both runs (apart from
their absolute values), while the extent of the regions characterised by
oxygen-rich gas is instead clearly larger in the \agn{} case than in the
\csf{} case.

From this comparison, we conclude that AGN feedback is more efficient
than stellar feedback in distributing the metal-rich gas out to large
cluster-centric distances, in small high-redshift haloes.  Since these
haloes and some of the expelled gas will be eventually accreted during
the cluster mass assembly, these processes essentially favour the
mixing of the pre-enriched gas that will end up in the main cluster at
$z=0$.

Therefore, AGN feedback also influences the distribution
of metal-rich gas at lower redshifts ($0 \lesssim z \lesssim 2$), which
is different in \agn{} and \csf{} cases.
This is quantitatively investigated through the radial profiles of
individual chemical species (Fe, Si and O), and abundance ratios (Si/Fe and O/Fe),
at redshifts $z=0,1,2$, in the following.

\subsection{Metallicity profiles}
\begin{figure}
\centering
\includegraphics[width=0.44\textwidth,trim=35 15 20 5,clip]{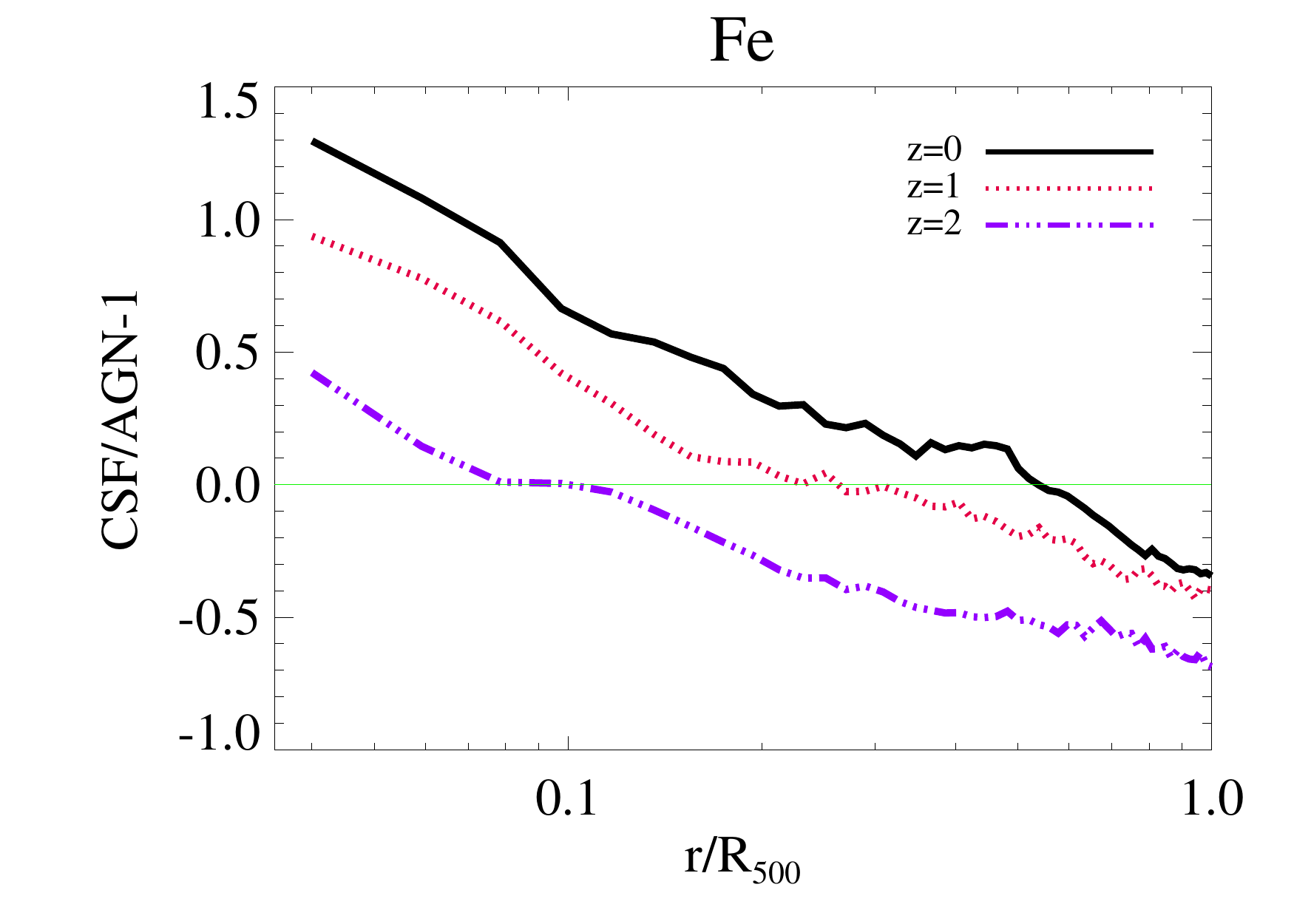}\\[10pt]
\includegraphics[width=0.44\textwidth,trim=40 15 20 5,clip]{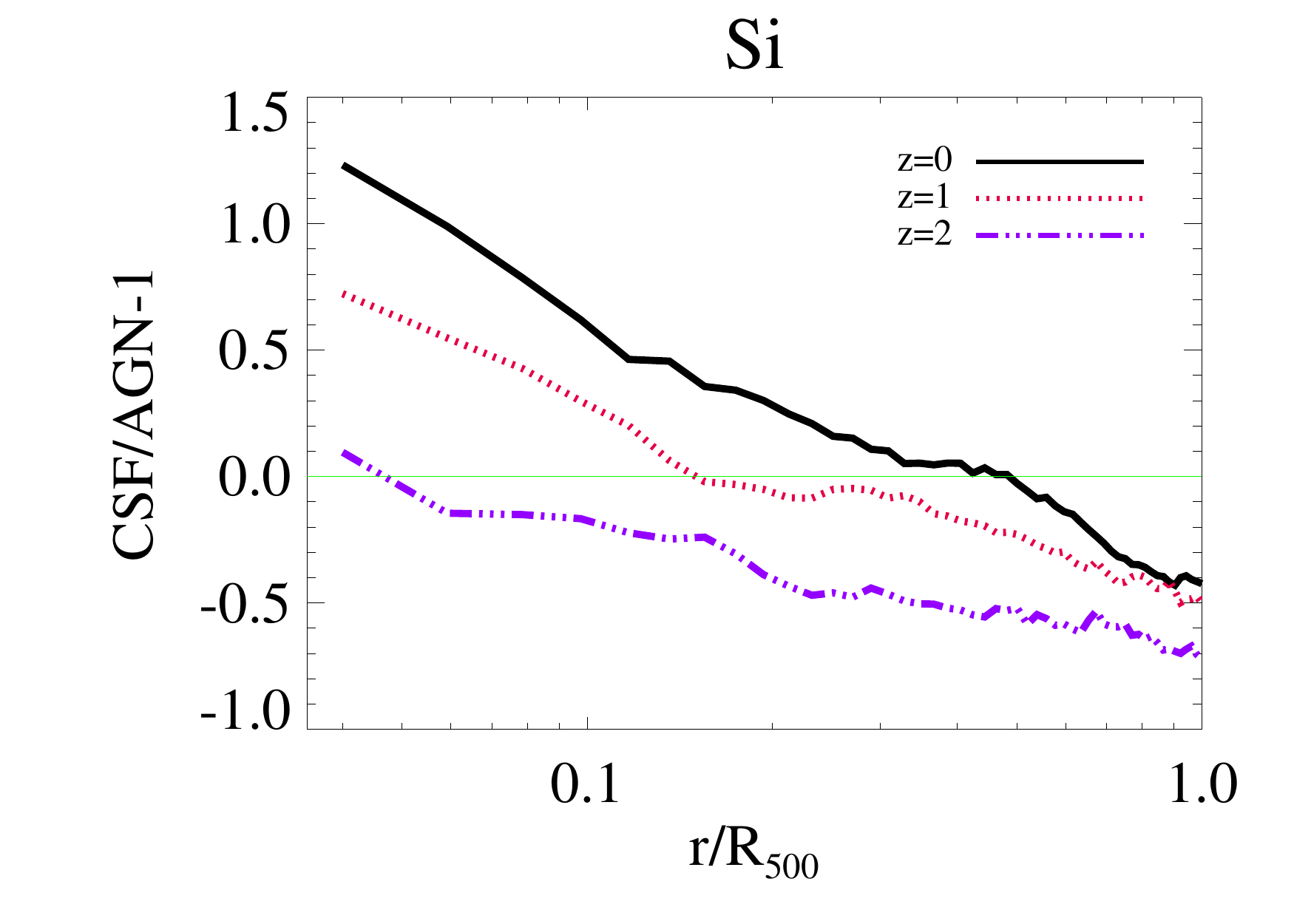}\\[10pt]
\includegraphics[width=0.44\textwidth,trim=40 15 20 5,clip]{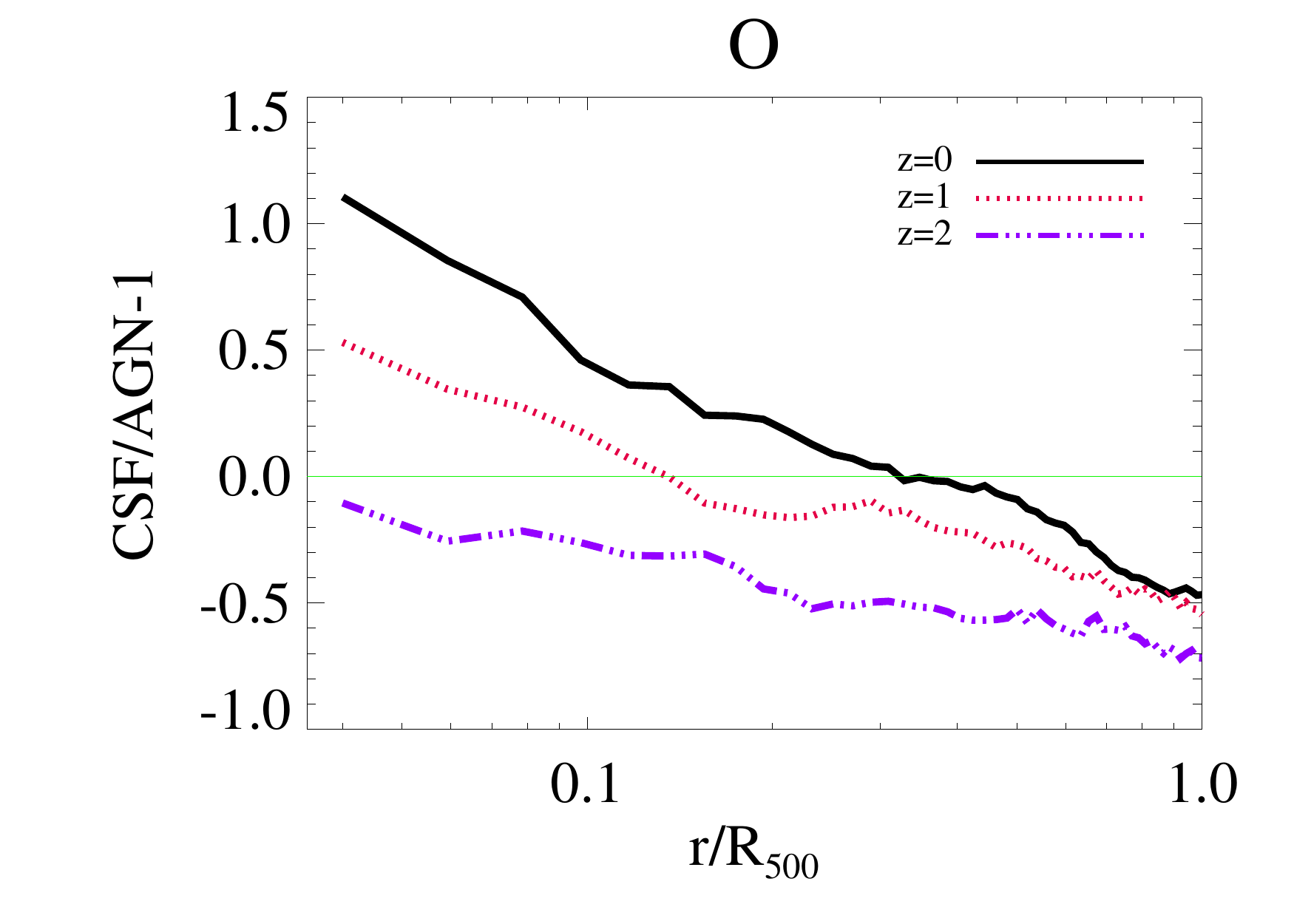}
\caption{Comparison between the median Fe, Si and O profiles in the \csf{} and \agn{} runs at $z=0,1,2$.}
\label{fig:csf-bh}
\end{figure}

In Fig.~\ref{fig:csf-bh}, we show the comparison between
  the iron, silicon and oxygen profiles in the \agn{} and \csf{} runs,
  by reporting the percentual difference between the two runs
  normalized to the \agn{} profile, versus the radial distance from
  the cluster centre.  This emphasizes the  difference between the
median abundance profiles in the \csf{} clusters with respect to the
run including AGN feedback, and their different evolution with
redshift.

At $z=0$ the \csf{} clusters typically have higher metal abundances than
\agn{} ones in the central part (by a factor of $\sim 2$), and lower metallicities in the outskirts (by $30$--$50\%$ at $\rfive$).
The region where the excess is noticed is different for the three chemical elements,
typically larger in the Fe case and smaller for Si and O,
and it increases in size going from $z=2$ to $z=0$.
This essentially indicates that in the centre of \csf{} clusters
the metal production continues till low redshift and contributes to
increase the average ICM metallicity
due to the lack of an efficient feedback mechanism,
able to hold back the overcooling of the gas, and quench star formation.
On the contrary, the negative values of the \csf{}-to-\agn{} profile
ratios in the outskirts indicate that gas metallicities at large
distances from the cluster centre are higher in the \agn{} simulations
than in the \csf{} run, because of the widespread distribution of
metal-enriched gas following the AGN-feedback activity at high
redshift ($z > 2$), as discussed with Figs.~\ref{fig:O-maps}
and~\ref{fig:SFR-maps} \cite[see also][for a study on the role
of AGN feedback on the distribution of gas at high redshift]{mccarthy2011}.

The redshift evolution of the \csf{}-to-\agn{} ratios from $z=2$ to
$z=0$ is also different for the three species: the iron \csf{}-to-\agn{}
curve is steep at all redshifts, while it gets from shallower
at $z=2$ to steeper at $z=0$ for Si and, especially, for O.
Therefore, the difference between the slopes of the Fe profiles in
the two runs must be already established at high redshift ($z=2$), while
it augments with time for both Si and O.  In
particular, the flatter (although still with a difference in the gradient $\gtrsim 2$)
 high-$z$ \csf{}-to-\agn{} ratios for the SNII
products indicate that, despite a different normalization, the
metallicity gradient in the two runs is more similar than it is at~$z=0$.

Differently from iron, the oxygen \csf{}-to-\agn{} ratio at $z=2$ is
negative at all radii, further indicating that at high redshift the
oxygen metallicity of \agn{} clusters is higher than in the \csf{}
simulation, also in more central regions.  The fact that SF should be partly
suppressed in the \agn{} run is nevertheless still consistent with
this picture.  Namely, SNII products are typically released in star
forming regions, and are therefore likely to be distributed among
star-forming gas particles.  Especially in the \csf{} case where SF is
not sufficiently compensated by any efficient feeback mechanism, this
means that elements like silicon and oxygen have a larger probability
to be rapidly locked back into newly formed stars, reducing therefore
the amount of available enriched gas.  Instead, given the longer
time-scale associated to iron production, mainly driven by SNIa, this
effect is expected to be less significant and the Fe abundance in the
core of \csf{} clusters still exceeds the one in the \agn{} case
\citep[see][]{Fabjan_2010}.  Going to lower redshifts, this different
behaviour of Fe with respect to O and Si is still visible from the
ratio of \csf{} to \agn{} profiles, which is greater for iron and
smaller for silicon and oxygen, in the innermost regions.  Despite
this, at recent times even the production of oxygen and silicon is
enhanced in the CSF runs due to the higher star formation with respect
to the AGN simulations, where the star formation has been quenched by
the powerful feedback mechanism.

\subsubsection*{Abundance ratios}

\begin{figure}
\centering
\includegraphics[width=0.45\textwidth,trim=40 15 20 5,clip]{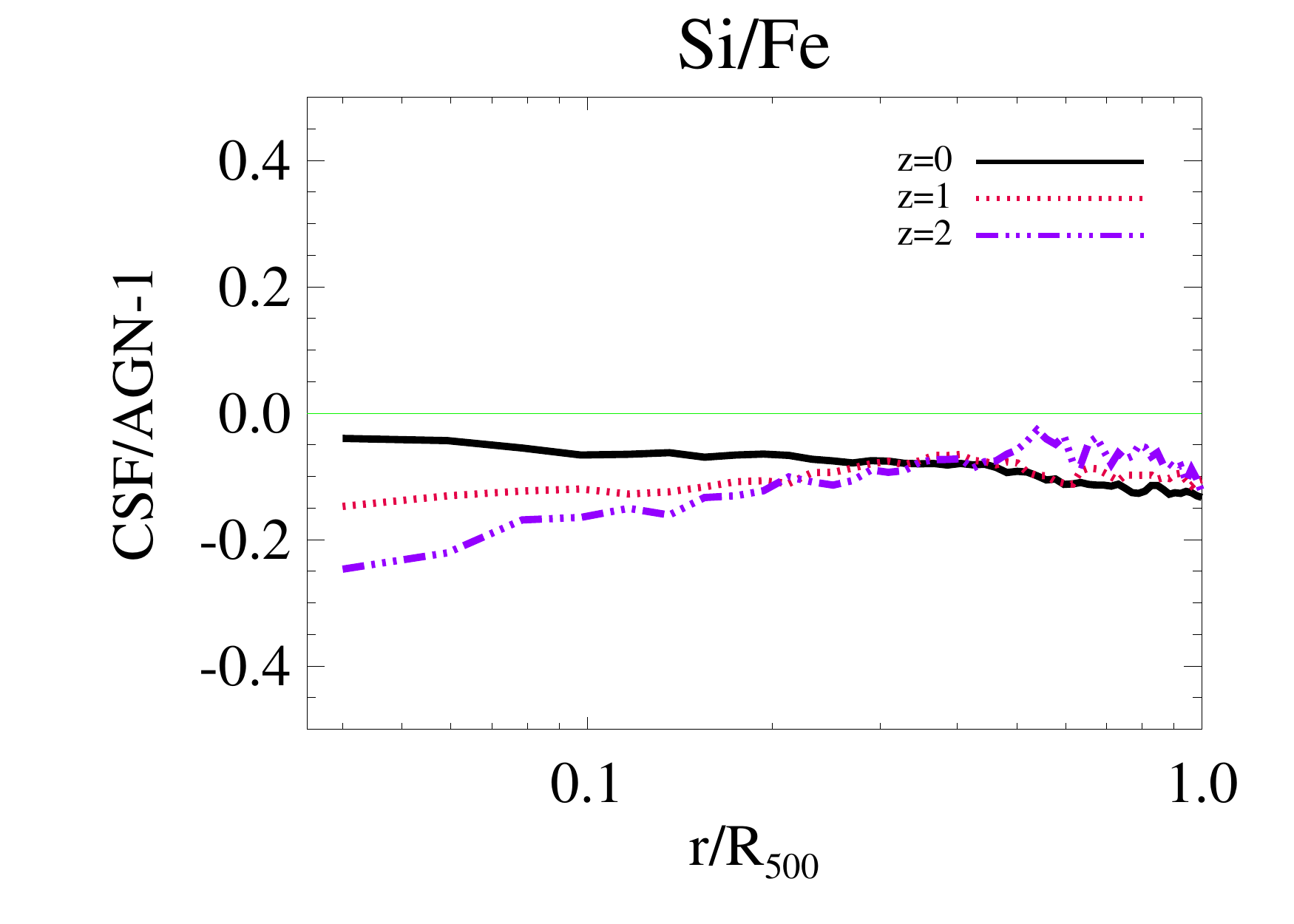}\\[15pt]
\includegraphics[width=0.45\textwidth,trim=40 15 20 5,clip]{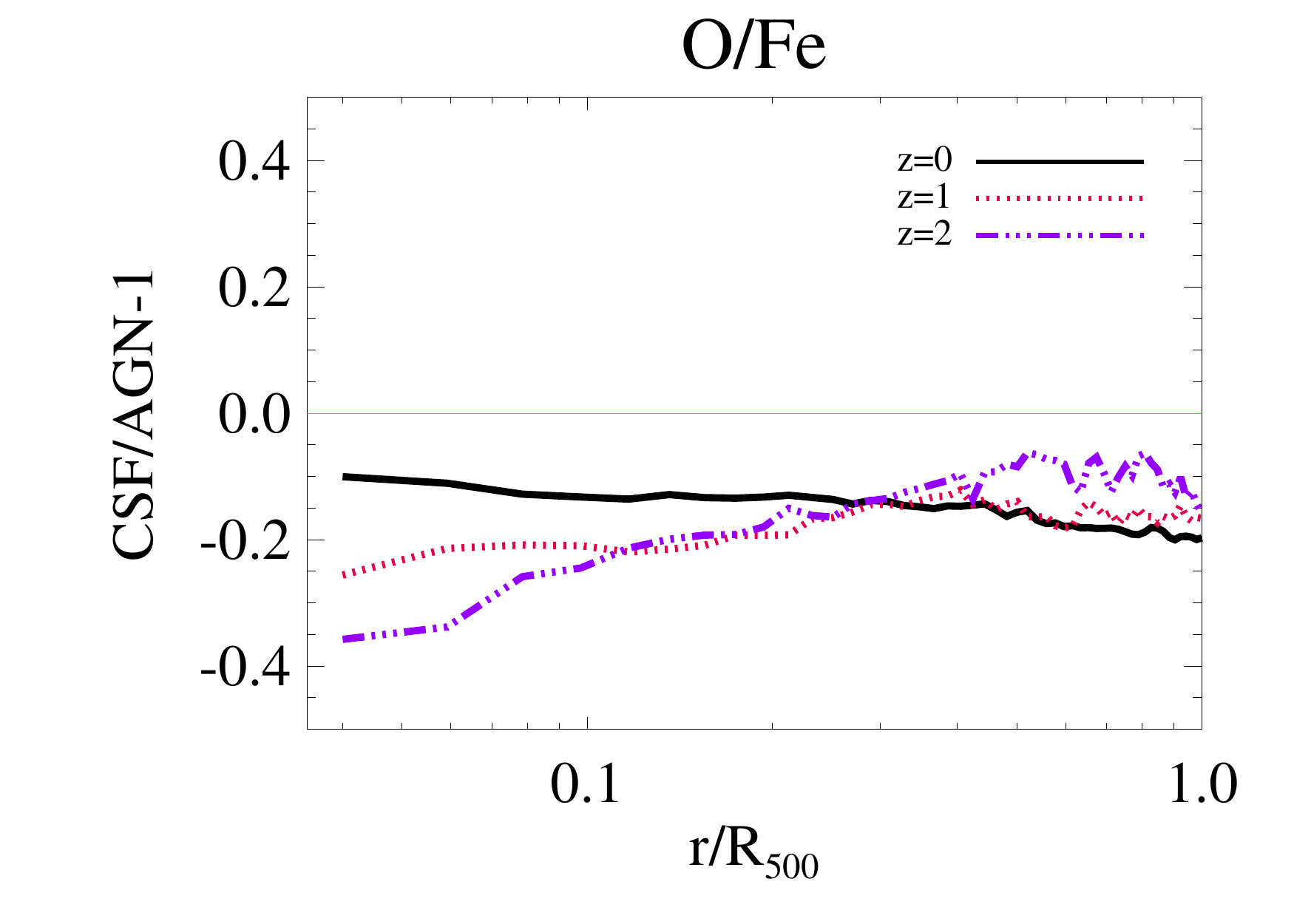}
\caption{Comparison between the median Si/Fe and O/Fe profiles in the \csf{} and \agn{} runs at $z=0,1,2$.}
\label{fig:csf-bh2}
\end{figure}

The behaviour of the three species discussed above
is reflected into the \csf{}-to-\agn{} comparison of
the abundance ratios and their evolution with redshift,
as shown in Fig.~\ref{fig:csf-bh2}.

In general, we see that the shape of the abundance ratio profiles in both
\agn{} and \csf{} runs are more similar at $z=0$ than at $z=2$.
At the present time, in fact, iron as well as silicon and oxygen
present similar differences between the two runs
(see Fig.~\ref{fig:csf-bh}).

From Fig.~\ref{fig:csf-bh2} (upper panel), the flat and negative
\csf{}-to-\agn{} curve for the Si/Fe ratio at $z=0$ indicates that the
Si/Fe profiles are almost parallel in the two runs, with an overall
lower normalization in the case of the \csf{} clusters.  The mild
decrease towards the outskirts (from $5\%$ in the centre out to $\sim 15\%$ at $\rfive$) also suggests that the difference is
not exactly constant over the radial range, but is enhanced at large
distances from the cluster centre.  This tells that SNII and SNIa
are overall distributed similarly in both runs, even though SNII
elements are less abundant, relative to Fe, in the  \csf{}
clusters (see Fig.~\ref{fig:csf-bh}).

On the contrary, at $z\sim 2$, the Si/Fe and O/Fe profiles are more
similar in the outskirts, whereas the picture of the innermost region
appears to be reversed. \csf{} clusters show a deficiency of SNII elements
with respect to iron in the central region which is more severe
than in the outskirts, because
Fe is more abundant in the core of \csf{} clusters than in the \agn{}
ones, while there is an opposite trend for Si and O (Fig.~\ref{fig:csf-bh}).

In general, we note that the difference between the \csf{} and the \agn{}
abundance ratios is always within $\sim 20\%$ in the region outside of
$\sim 0.1$--$0.15\,\rfive$.
Only in the central part, instead, there can be up to $\sim 40\%$
difference in the O/Fe profile at $z=2$.

\section{Tracing the contribution from different enrichment sources}\label{sec:track}
We investigate
in this section the effective contributions to the ICM metal
enrichment by the three stellar sources explicitely traced in the
simulations, namely SNIa, SNII and AGB stars.
As a study case, we restrict the analysis to four representative
clusters, selected at $z=0$ to be two massive haloes and two smaller
systems, and for each pair of clusters we choose one CC and one NCC cluster.%
\begin{figure*}
\centering
{\Large  {\bf CC}: \qquad\qquad\qquad D2 \hfill D10 \hphantom{\Large  CC: \qquad\qquad\qquad}}\\
\includegraphics[width=0.49\textwidth]{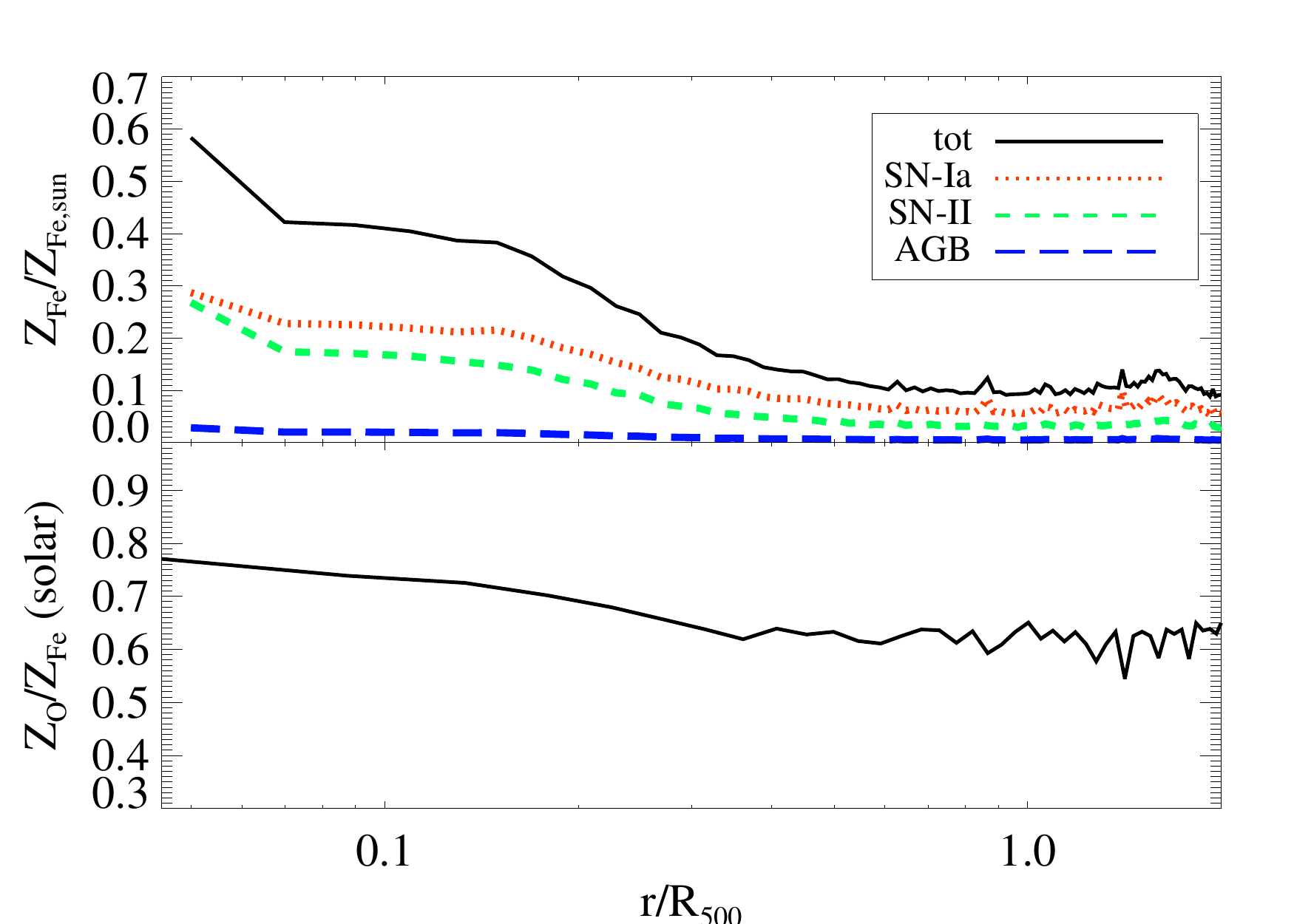}
\includegraphics[width=0.49\textwidth]{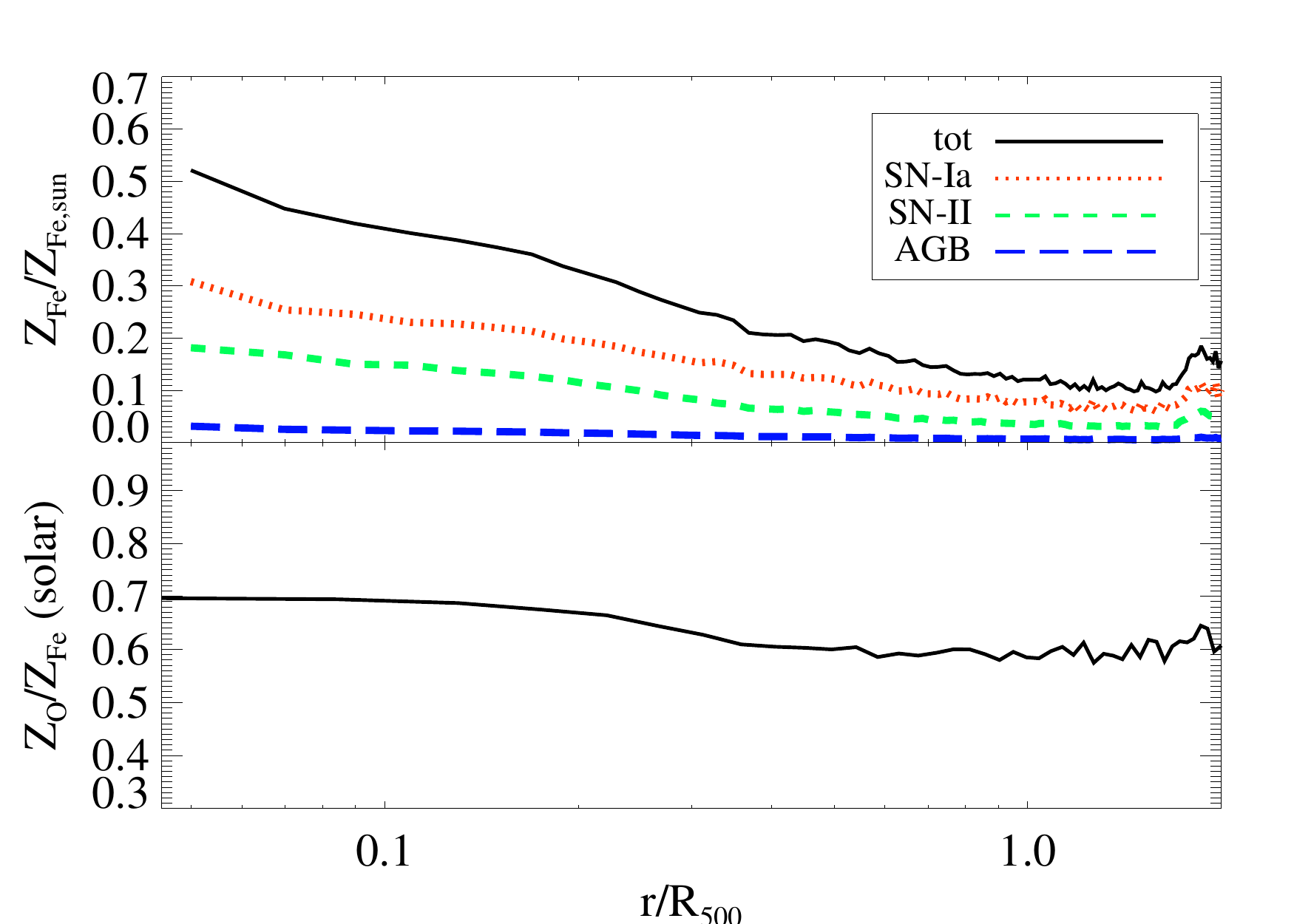}\\[10pt]
{\Large  {\bf NCC}: \qquad\qquad\qquad D3 \hfill D6 \hphantom{\Large  NCC: \qquad\qquad\qquad}}\\
\includegraphics[width=0.49\textwidth]{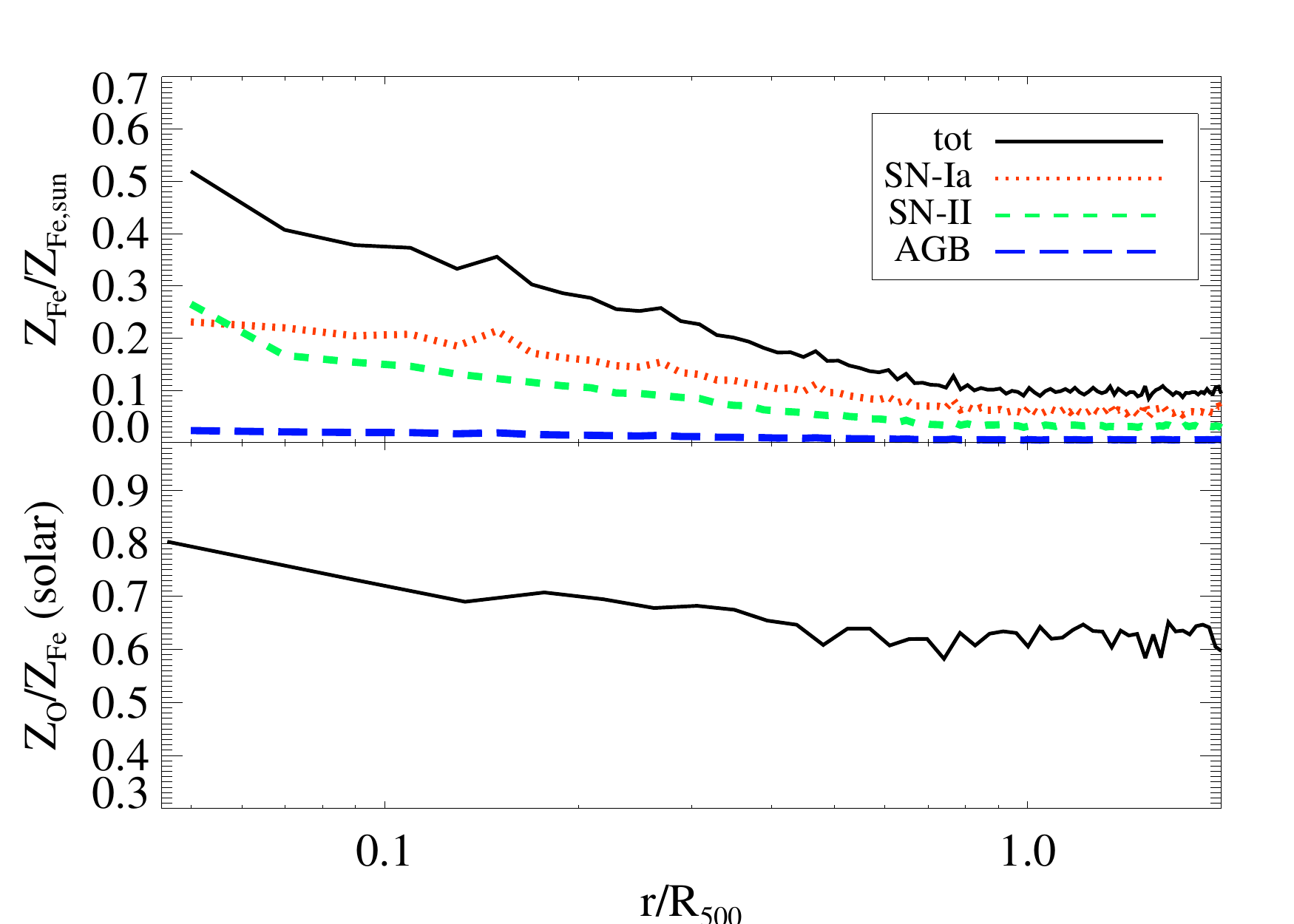}
\includegraphics[width=0.49\textwidth]{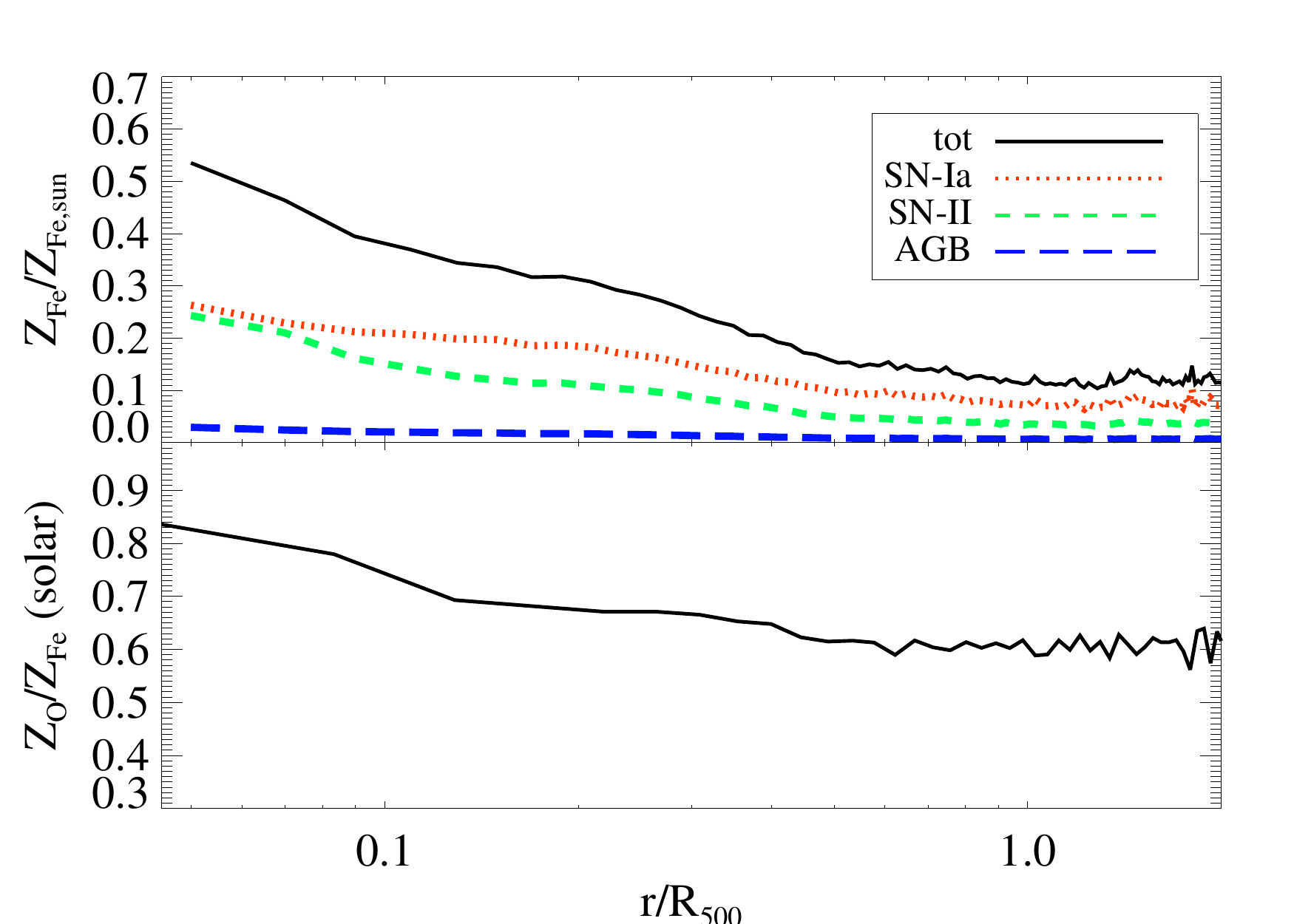}
\caption{Upper insets: mass-weighted radial profiles of Fe abundance at $z=0$,
  differentiating the different enrichment sources: SNIa (magent
  dotted line), SNII (blue, dashed line) and AGB (green long-dashed
  line). For comparison, also the total iron-abundance profile is
  shown (black, solid line).  Lower insets: mass-weighted radial profile of the
  observable tracer, O/Fe, of the relative contribution between SNII
  and SNIa products.  In the upper row the clusters are CC systems
  (D2 and D10, from left to right respectively), while the two
  clusters in the lower row are NCC (D3 and
  D6).\label{fig:enrich-profs2}}
\end{figure*}

Considering iron as a typical tracer of ICM metallicity,
we explore the $z=0$ Fe abundance profile cluster by cluster, computed
for the three aforementioned enrichment sources separately.
This is shown in Fig.~\ref{fig:enrich-profs2} (upper inset in each
panel).
From the comparison against the oxygen-to-iron abundance ratio,
reported in the lower inset of each panel,
we can investigate directly
the capability of this observable quantity to unveil the relative
contribution of SNII and SNIa to the ICM enrichment.

From the Fe profiles in Fig.~\ref{fig:enrich-profs2}, we confirm that
AGB stars basically release an almost null amount of Fe, except for
the contribution due to the iron locked within the stars that form
from gas that was previously enriched with it.  Inspecting the four sets of
profiles, we also see that there is a general tendency for the SNIa
and SNII Fe-profiles to be similarly flat in the cluster outskirts.
This resemblance is the origin of the flatness of the O/Fe profile in the outer
region of the clusters, pointing to a similarly homogeneous distribution
of the metal products from the two enrichment channels.
Towards the central regions, instead, the iron abundance profile
corresponding to SNII enrichment typically presents a steepening with
respect to the SNIa one, which is overall higher but flatter
throughout the entire radial range. This trend is mirrored
by the typical increase of the O/Fe abundance ratio in the
cluster innermost~region.


\begin{figure}
\centering
\includegraphics[width=0.47\textwidth]{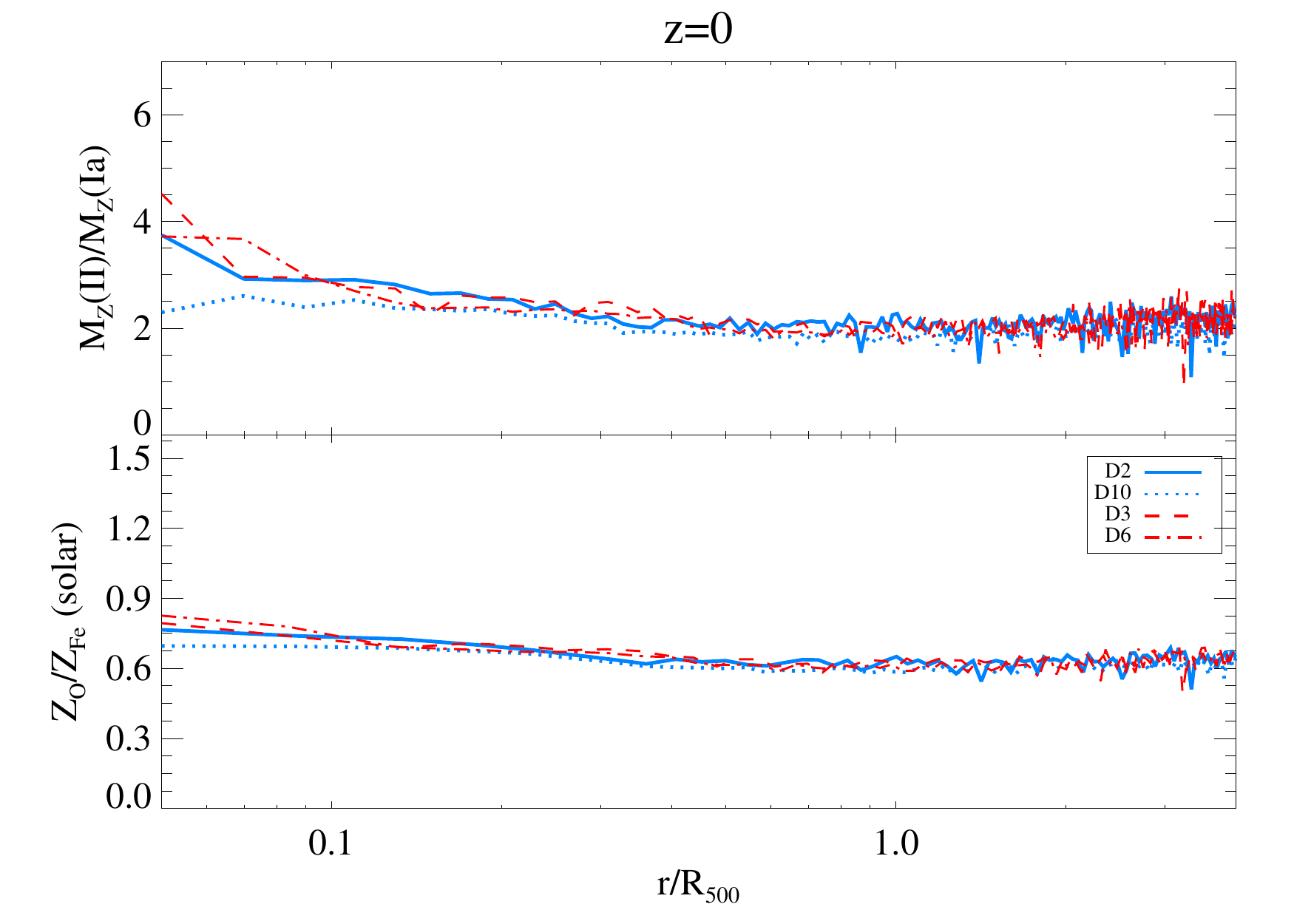}
\includegraphics[width=0.47\textwidth]{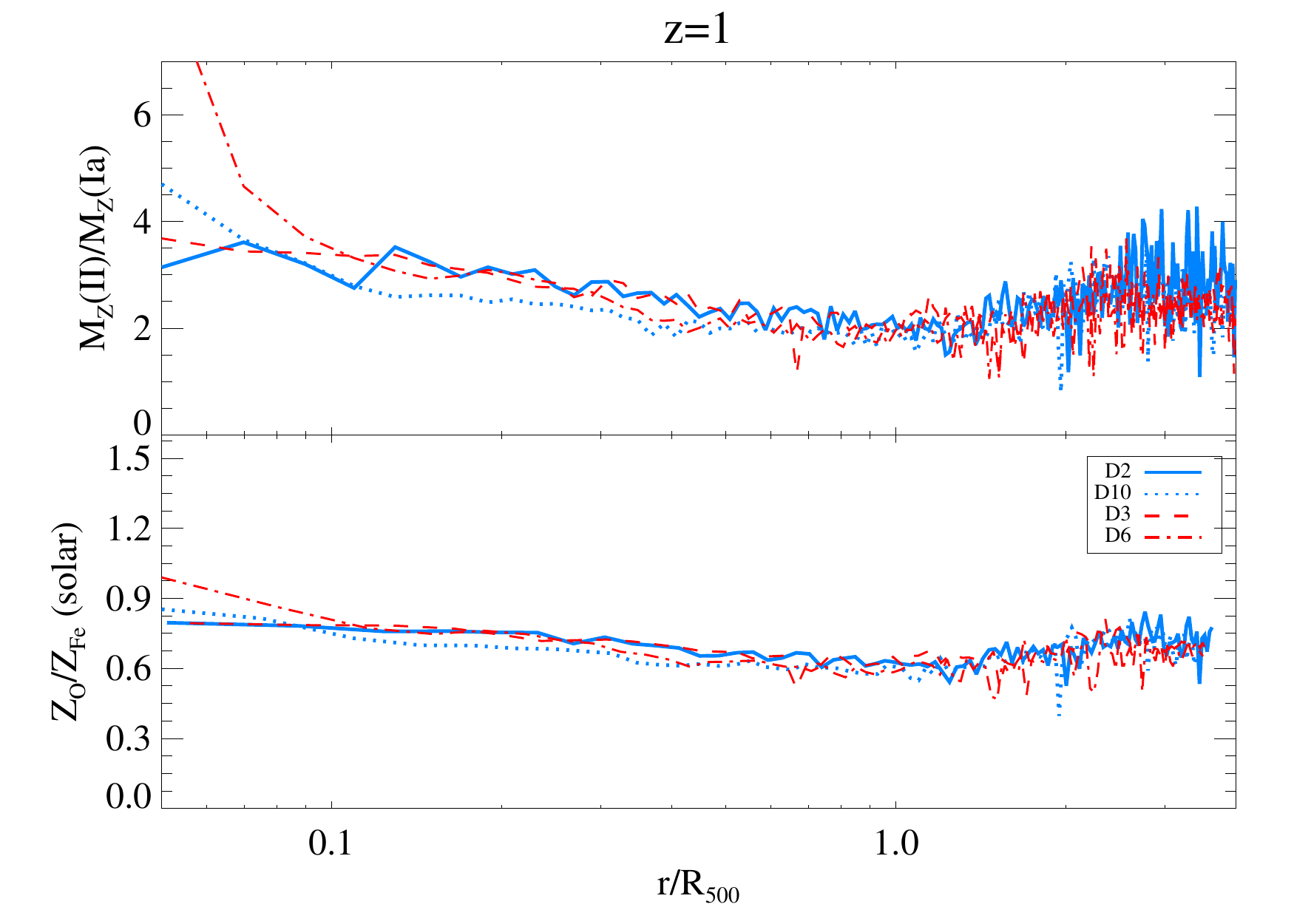}
\includegraphics[width=0.47\textwidth]{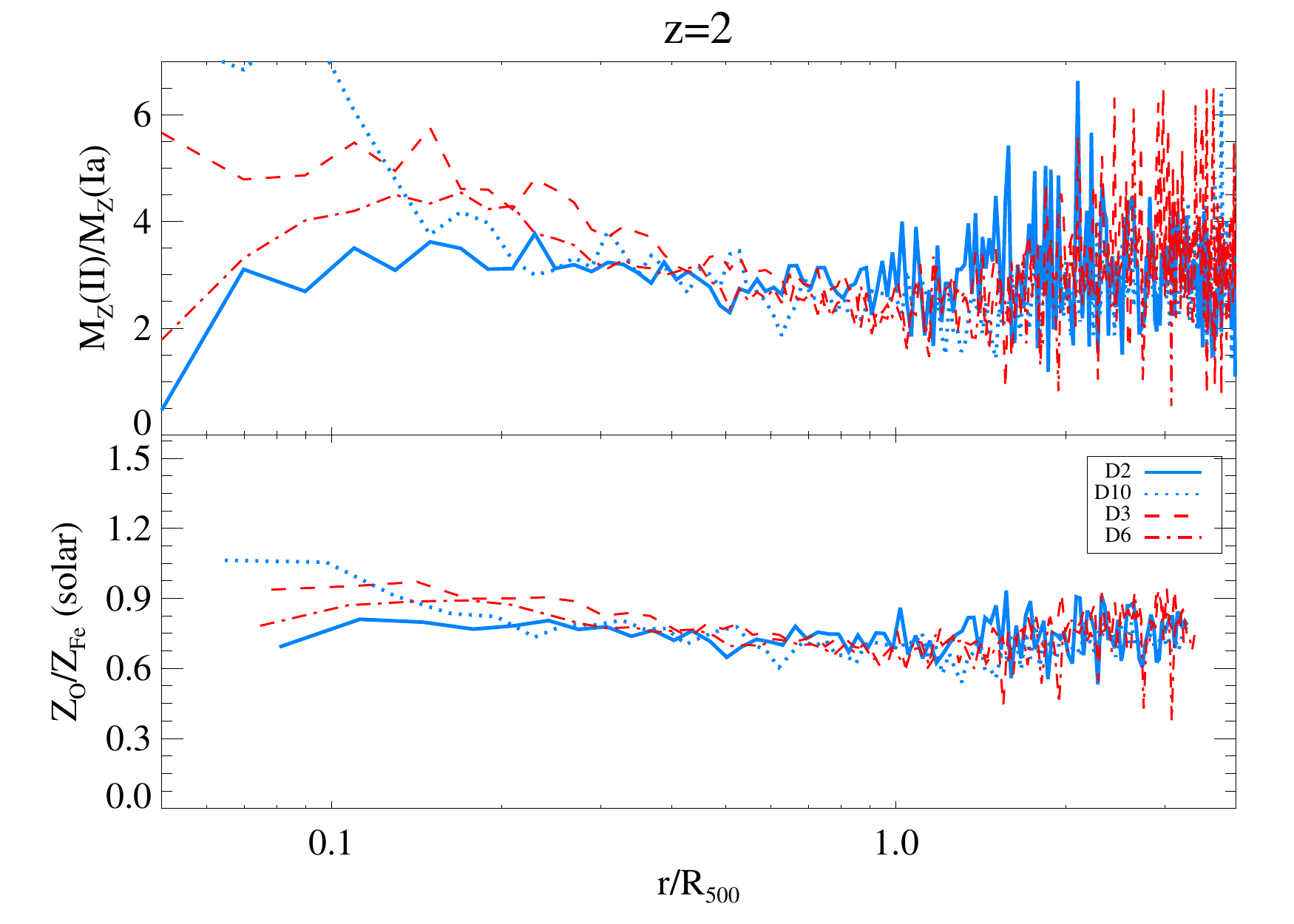}
\caption{In each panel, we show the comparison of (i) the ratio
  between the metals released by SNII and SNIa (top);
  and (ii) the mass-weighted radial profile of the O/Fe abundance ratio.
  From top to bottom,
  the redshifts shown are $z=0$, $z=1$ and $z=2$, respectively. As in
  the legend, the different lines refer to the four clusters
  considered, with blue and red marking the CC and NCC classification
  done at $z=0$.\label{fig:enrich-profs}}
\end{figure}

Taking advantage of the information in the simulations,
we can actually directly compute the contribution from
SNII and SNIa to the total ICM metal content.
In Fig.~\ref{fig:enrich-profs}, we present in the upper inset of each
panel the ratio between the mass of metals produced by SNII and
that produced by SNIa, $M_Z(II)/M_Z(Ia)$, as a function of radius.
This is compared as well to the observational O/Fe~tracer.

At $z=0$, the shape of the two profiles is very
similar and both indicators agree in a flat relative distribution of
SNII to SNIa metal enrichment outside $0.1$--$0.2\,\rfive$.
The main difference between the two indicators is given by the larger
scatter in the $M_Z(II)/M_Z(Ia)$ ratio, especially in the central part
and in the outermost regions.  This is mainly due to the difference in
the actual relative contributions probed: the iron metallicity
considered within the O/Fe abundance ratio is in fact contributed by
both SNII and SNIa, even though with different proportions (as seen in
Fig.~\ref{fig:enrich-profs2}).  We verified that the profiles of
$M_Z(II)/(M_Z(II)+M_Z(Ia))$ present a significantly lower scatter,
and essentially show a strong similarity with the O/Fe profiles.
By examining the evolution of these profiles, shown in the central and
bottom panels, we note that there is no particular trend associated to
the classification of the clusters as CC or NCC at $z=0$.

Independently of the enhancement previously discussed for the
$M_Z(II)/M_Z(Ia)$ case, we find that the scatter among the different
clusters augments in both indicators at higher redshift, especially
within $0.3$--$0.4\,\rfive$, where differences between the specific
star formation history of the clusters become more important.
At large distances from the cluster centre the profiles
tend to become also noisier, which is likely due to the
more clumpy distribution of the metals at early times with respect to
the more homogeneous one at $z=0$, when mixing processes of the gas
have had more time to be effective.
At all redshifts, part of the noise that characterises the region
outside $\rfive$ remains because those outer regions are more
prominently affected by the presence and the infall of smaller
substructures.

Overall, we find that in the cluster outskirts (outside of
$0.3$--$0.4\,\rfive$) the relative contribution of SNII and SNIa to
the enrichment does not change significantly from $z=2$ to $z=0$,
except for a further flattening of the profiles and a decrease of the
scatter due to the progressive contribution of late-time enrichment by
SNIa.


\section{Summary and conclusion}\label{sec:conclusions}

In this paper we have presented an analysis of the ICM chemical
properties for a sample of 29 clusters from zoom-in, cosmological,
hydrodynamical simulations.
As a first step, we compare the results at $z=0$ with
observational findings available from the literature, obtained with
X-ray telescopes such as \chandra, \xmm\ and \suzaku.
We subsequently explore the role of different feedback sources and
enrichment channels on the distribution of the metal-rich ICM at $z=0$
and at different epochs through the cosmic time, by comparing
two runs of the simulations, with and without the treatment of AGN
feedback (\agn{} and \csf{} runs, respectively).

The main results of this work are summarized in the following.

\begin{itemize}
\item In our set of simulations, able to produce the
  observed diversity of central properties that discriminate between
  CC and NCC systems, we also recover the observed anti-correlation
  between entropy and metallicity in cluster cores at $z=0$.
  Considering the intrinsic (mass-weighted and three-dimensional) quantities, such anti-correlation is instead weaker than
  observed and tends to fade with increasing redshift.
\item We find overall good agreement between radial profiles of ICM
  metallicity and abundance ratios predicted by the simulations and
  available observational data at $z=0$.
\item We find relatively flat abundance ratio profiles in cluster
  outskirts ($r \gtrsim 0.1$--$0.2\,R_{180}$) out to very large
  cluster-centric distances, both in the emission-weighted projected
  profiles and in the intrinsic (mass-weighted, 3D) ones.
\item The level of enrichment in the intermediate and outer
  regions is already established at $z\sim 1$--$2$ and does not
  depend on the cool-coreness of the halo at any times.
  \item Overall, we find no evidence for significant metallicity
    evolution of the hot gas in the whole sample, between $z=2$ and
    $z=0$; instead, we find an increase of the metal mass
     in both gas and stars, with an increasing
    metallicity of the star-forming gas component, in the innermost
    region.
\item The homogeneous distribution out to large radii of SNII
  products (e.g. oxygen), typically associated to star-forming
  regions, is mainly driven by the high-redshift AGN feedback, which
  favours the spatial distribution and mixing of already enriched gas
  in the small substructures that eventually form the cluster.
\end{itemize}

  Our findings strongly support the idea that the wide spreading of
  metal-enriched gas, which turns into the remarkably homogeneous
  enrichment found in the outskirts of low-redshift clusters, has
  happened mostly
  at early times ($z>2$--$3$), during --- but even before
  --- the epoch of maximal star-formation and AGN activity~\cite[see also recent results by][]{ezer2016}.  In
  particular, we found that the early AGN feedback acting on
  high-redshift ($z > 2$) small haloes, characterized by shallower
  potential wells, is particularly efficient in spreading the
  metal-enriched gas out to the halo virial radius and well beyond.
  This is the crucial step to distribute the enriched gas over large
  regions, even before the merging of the sub-structures onto the main
  cluster. This explains the homogeneous metallicity distribution
  observed at large cluster-centric distances at $z\lesssim 2$ down to
  $z=0$, even in the case of SNII elements like oxygen, which is
  typically associated to recent star-formation episodes.
  In this framework, the flat proportion between SNII and
  SNIa products further indicates that both populations of SN
  contributed to that early enrichment, and their relative
  contribution to the total mass of metals produced does not change
  significantly from $z=2$ to $z=0$.

\looseness=-1
  These results essentially confirm the observational findings
  obtained so far by exploring the outer volume ($\rfive \lesssim r
  \lesssim\rtwo$) of massive clusters in the local
  Universe~\cite[e.g. with \suzaku{};][]{werner2013,simionescu2015}.
  Nevertheless, due to their low surface brightness, cluster outskirts
  still remain a largely un-explored territory for studies aimed at
  precisely estimating heavy elements abundances, which are on the
  contrary much better constrained in the innermost
  regions~\cite[e.g.][]{dePlaa_2007,tamura2009,de_Plaa_2013,mernier2016a,mernier2016b}.
Beyond the innermost regions, in fact, observational data are mostly
limited in their statistical quality, which makes it very difficult to
put tight constraints on the ICM metallicity.  The thermodynamical
structure and multi-phaseness of the ICM at large radii, typically
beyond $0.2$--$0.4\,R_{180}$, is poorly known and leaves us with
substantial uncertainties about a major portion of the ICM content. In
fact, at intermediate and large cluster radii observations can be
strongly contaminated by the background~\cite[see][for a dedicated
  study]{Molendi_2016}.
  In order to place tighter constraints on metal abundances in the
  periphery of clusters, next-generation X-ray missions with high
  spectral resolution and large effective area, such as
  \athena{}\footnote{\tt
    http://www.the-athena-x-ray-observatory.eu}~\cite[][]{athena}, are
  definitely needed. This will improve substantially our understanding
  of the origin of the metal-rich gas~\cite[][]{athenaettori2013}, by
  investigating the regions where the accretion and mixing processes
  are still ongoing.  Also, such observational advancements will allow
  us to push the study of the ICM chemical properties out to higher
  redshifts~\cite[][]{athenaPointecouteau2013}, closer to the epoch
  when the bulk of the enrichment happened.

\enlargethispage*{\baselineskip}
  From the numerical point of view, state-of-the-art simulations, like
  those employed for the present analysis, managed to successfully
  model, for the first time, the thermal properties as well as the
  radial distribution of the heavy elements in the ICM, from the core
  region out to the boundary of low-redshift massive clusters.
  Nevertheless, we also found interesting discrepancies between
  predictions from simulations and observations, for instance
  concerning the evolution of the typical metallicity within the core
  of the clusters. To further explore this issue, simulations with
  higher resolution and larger cluster samples are required, so that
  the relation between central entropy and metallicity can be investigated
  for statistically significant samples of CC and NCC, up to high~redshifts.
  Furthermore, the possibility to track the origin of the metals in
  the ICM, both spatially and in terms of enrichment sources
  (e.g. SNII, SNIa, AGB stars), can pave the way for a number of
  focused studies: on the relation between specific feedback sources
  and enrichment channels; on the origin in space and time of the gas
  enriched by a specific stellar source and localized in the
  outskirts of clusters at $z=0$; on the dynamical mixing of specific
  metal species depending on the sources that produced them.

  As a final remark, we note that additional mechanisms connected to
  the interplay between BH accretion and ICM properties (such as
  kinetic feedback and inflation of bubbles of gas due to jets
  associated to the AGN, or possible effects of magnetic fields) as
  well as metal diffusion
  and depletion of heavy elements due to dust formation,
  that are still not included in our simulations, might play an
  important role in shaping the distribution of the metal-rich gas,
  especially in the outskirts.

\section*{Acknowledgements}
The authors would like to thank S.~Molendi, A.~Simionescu and
N.~Werner for useful discussions, V.~Springel for allowing us to
access the developer version of the {\small GADGET} code and
L.~Steinborn for providing access to the improved AGN feedback model.
Also, we thank the anonymous referee for constructive comments on the
paper that helped improving the presentation of our results.  We
acknowledge financial support from PIIF-GA- 2013-627474, NSF
AST-1210973, PRIN-MIUR 201278X4FL, PRIN-INAF 2012 ``The Universe in a
Box: Multi-scale Simulations of Cosmic Structures'', the INFN INDARK
grant, ``Consorzio per la Fisica'' of Trieste, CONICET-Argentina,
FonCyT.  Simulations have been carried out using Flux HCP Cluster at
the University of Michigan, Galileo at CINECA (Italy), with CPU time
assigned through ISCRA proposals and an agreement with the University
of Trieste.  We also acknowledge PRACE for awarding us access to
resource ARIS based in Greece at GRNET, through the DECI-13 PRACE
proposal.  The post-processing has been performed using the PICO HPC
cluster at CINECA through our expression of interest.  SP also
acknowledges support by the {\it Spanish Ministerio de Econom{\'i}a y
  Competitividad} (MINECO, grant AYA2013-48226-C3-2-P) and the
Generalitat Valenciana (grant GVACOMP2015-227).  DF acknowledges
partial support by the Slovenian Research Agency.  MG is supported by
NASA through Einstein Postdoctoral Fellowship Award Number PF-160137
issued by the Chandra X-ray Observatory Center, which is operated by
the SAO for and on behalf of NASA under contract NAS8-03060.
\bibliographystyle{mnbst}
\bibliography{metals.bib}


\appendix

\section{CC in the CSF simulation?}
Despite the majority of the clusters in the \csf{} run appear to be
NCC clusters, as well known from previous analyses, we note from the
bottom panel in Fig.~\ref{fig:entr-met0} that 5 clusters of our \csf{}
sample have 2D pseudo-entropy below the $\sigma=0.55$ threshold, and
therefore could be classified as CC systems.  We inspected these 5
clusters in more detail.  Out of these, 2 clusters --- those close to
the threshold line --- are not confirmed to be real CC systems when
the value of the pseudo-entropy $\sigma$ is computed intrinsically for
the three-dimensional spherical ``IN'' region.
This is also confirmed by the visual
inspection of their entropy profiles and maps.

\begin{figure}
\centering
\includegraphics[width=0.48\textwidth]{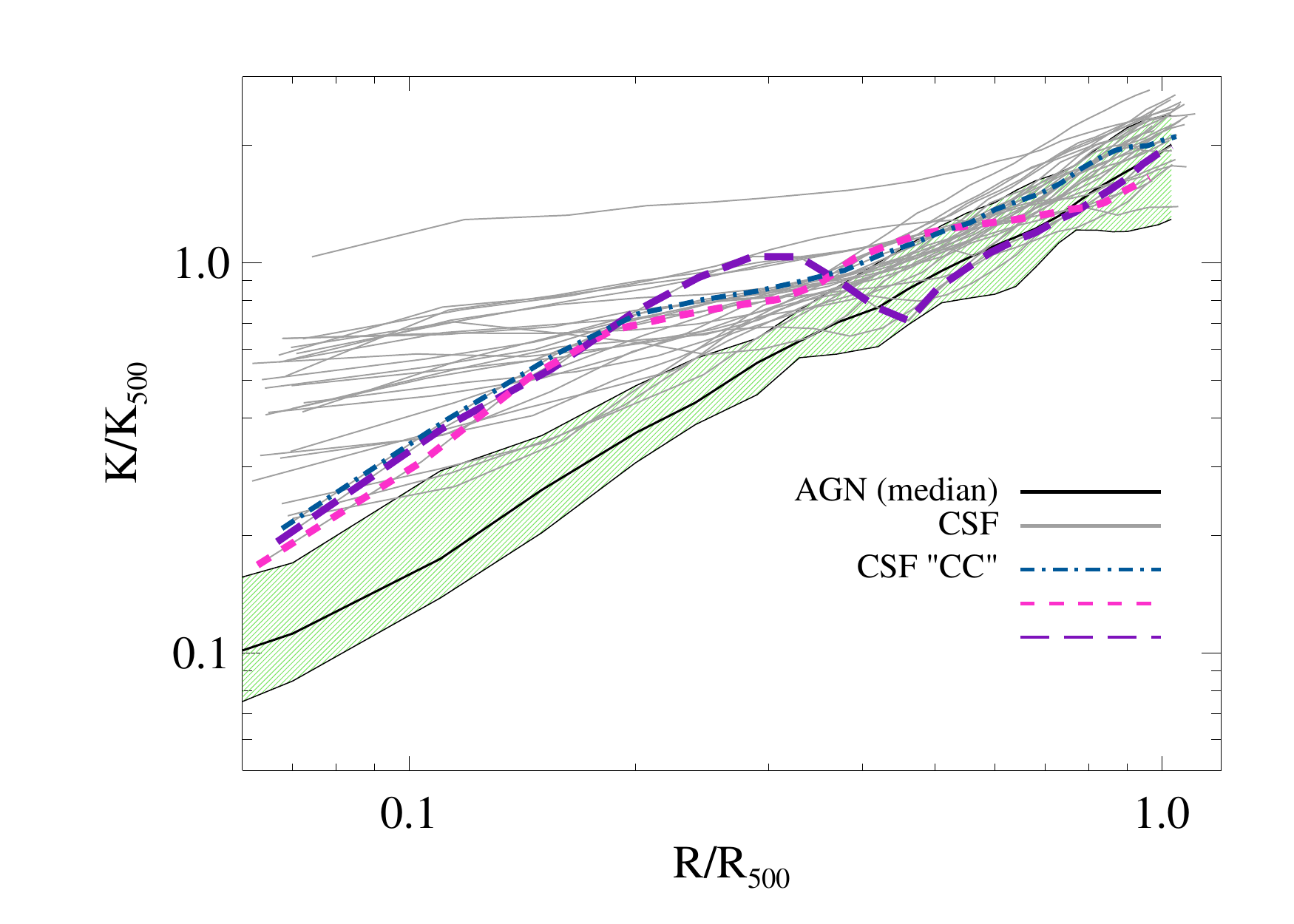}
\caption{Comparison between the median entropy profile of the
  \agn{} CC clusters (black solid line and shaded area) and the single
  entropy profiles for the \csf{} sample (grey thin lines). Marked in coloured thick
  lines are the 3 clusters with
  $\sigma<0.55$.}\label{fig:csf-entr-prof}
\end{figure}

The 3 remaining clusters with the lowest values of $\sigma$ in the
projected emission-weighted case, present a value of the
pseudo-entropy in the core region that is lower than $0.55$ even when
the 3D mass-weighted estimate is considered.  Nevertheless, these
clusters show irregular entropy profiles, that are generally higher
and less smooth than the typical profiles of the CC systems obtained
in the \agn{} sample, which are instead in good agreement with
observations~\cite[]{rasia2015}.  The three objects are highlighted in
Fig.~\ref{fig:csf-entr-prof}, where the other individual entropy
profiles of the \csf{} clusters are marked in thin grey lines and the
median entropy profile of CC systems in the \agn{} sample is also
reported (black line with shaded area).

%
%


\end{document}